\documentclass{elsart}
\usepackage{graphicx}% Include figure files
\usepackage{dcolumn}% Align table columns on decimal point
\usepackage{bm}% bold math
\usepackage{color}
\usepackage{latexsym}
\journal{Nuclear Physics A}
\usepackage{latexsym}
\begin{document}
\newcommand{\leftg}{\langle \phi_0 |}
\newcommand{\rightg}{| \phi_0 \rangle}
\newcommand{\chiral}{\langle \bar{q} q \rangle} 
\newcommand{\vs}{\vspace{-0.25cm}}
\newcommand{\gev}{\,\mathrm{GeV}}
\newcommand{\mev}{\,\mathrm{MeV}}
\newcommand{\fmd}{\,\mathrm{fm}^{-3}}
\newcommand{\ord}[1]{\mathcal{O}(k_f^{#1})}
\newcommand{\dif}{\mathrm{d}}
\begin{frontmatter}
\title{Relativistic nuclear energy density functional  
constrained by low-energy QCD\thanksref{now}}
\thanks[now]{Work supported
in part by BMBF, DFG, GSI and MURST.}
\author[Mun]{P. Finelli},
\author[Mun]{N. Kaiser},
\author[Vre]{D. Vretenar},
\author[Mun]{W. Weise}
\address[Mun]{Physik-Department, Technische Universit\"at M\"unchen,
D-85747 Garching, Germany}                                                    
\address[Vre]{Physics Department, Faculty of Science, University of
Zagreb, 10 000 Zagreb, Croatia}

\begin{abstract}
A relativistic nuclear energy density functional is developed, 
guided by two important 
features that establish
connections with chiral dynamics and the symmetry breaking pattern 
of low-energy QCD:
a) strong scalar and vector fields related to in-medium changes of 
QCD vacuum condensates; 
b) the long- and intermediate-range interactions generated by one-and 
two-pion exchange, 
derived from in-medium chiral perturbation theory, with explicit 
inclusion of $\Delta(1232)$ 
excitations. Applications are presented for binding energies, radii 
of proton and neutron 
distributions and other observables over a wide range of spherical 
and deformed nuclei from 
$^{16}O$ to $^{210}Po$. Isotopic chains of $Sn$ and $Pb$ nuclei are 
studied as test cases 
for the isospin dependence of the underlying interactions. The 
results are at the same level 
of quantitative comparison with data as the best phenomenological 
relativistic mean-field 
models.
\end{abstract}  
\date{\today}
\end{frontmatter}

\section{\label{secI}Introduction.}
The successes of modern nuclear structure models 
in predicting many new phenomena in regions of 
exotic nuclei far from stability, 
and the recent applications of chiral effective
field theory to nucleon-nucleon scattering and the few-body problem,
have highlighted one of the fundamental problems in
theoretical nuclear physics: the relationship
between low-energy, non-perturbative QCD and the rich structure 
of nuclear many-body systems.

The most complete and accurate description of structure phenomena 
in heavy nuclei is currently 
provided by self-consistent 
non-relativistic and relativistic mean-field approaches. 
By employing phenomenological effective interactions,
adjusted to empirical properties of symmetric and asymmetric 
nuclear matter, and to bulk properties of stable nuclei,  
self-consistent mean-field methods have achieved a high level of 
precision in describing ground states and
properties of excited states in stable nuclei, exotic nuclei
far from $\beta$-stability, and in nuclear systems at the nucleon
drip-lines \cite{BHR.03,VALR.05}.

The self-consistent mean-field approach to nuclear structure
represents an approximate implementation of Kohn-Sham density 
functional theory (DFT) \cite{HK.64,KS.65,Kohn_Nobel,Dr.Gro}. The DFT
provides a description of the nuclear many-body problem in terms of
an energy density functional, $E[\rho]$. Mean-field models
approximate the exact energy functional which includes all
higher-order correlations. A major goal of nuclear structure theory is 
to build an energy density functional which is universal \cite{HK.64},
in the sense that the same functional is used for all
nuclei, with the same set of parameters. 
This framework should then provide a reliable microscopic description of 
infinite nuclear and neutron matter, ground-state properties of 
bound nuclei, rotational spectra, low-energy vibrations and large-amplitude 
adiabatic properties \cite{LNP.641}.   

In order to formulate a microscopic nuclear energy density functional, 
one must be able to go beyond the mean-field approximation 
and systematically calculate the exchange-correlation part, 
$E_{\rm xc}[\rho]$, of the 
energy functional, starting from the relevant
active degrees of freedom at low energy. 
The {\em exact} $E_{\rm xc}$ includes all 
many-body effects. Thus the usefulness of DFT crucially depends 
on our ability to construct accurate approximations to the 
exact exchange-correlation energy. The natural microscopic framework is 
chiral effective field theory. It is based on the separation of scales 
between long-range pion-nucleon dynamics, described explicitly, 
and short-distance interactions not resolved in detail at low energies.

An extensive program, synthesizing effective field theory methods and density 
functional theory, has recently been initiated by 
Furnstahl and collaborators \cite{PBF.03,BF.05,Fur.04}. 
For a dilute, confined 
Fermi system with short-range interactions, an effective action 
formalism leads to a Kohn-Sham density functional by applying 
an inversion method 
order-by-order in the relevant small expansion parameter, the local 
Fermi momentum times the scattering length. The starting point is 
a generating functional with external sources coupled to composite 
Fermion operators, e.g. particle number densities and 
kinetic energy densities. A functional
Legendre transformation with respect to source fields leads to 
an effective action functional from which the energy density functional 
is calculated. 

An alternative approach to the nuclear energy density functional, 
emphasizing links 
with low-energy QCD and its symmetry breaking pattern, has recently been 
introduced \cite{Fi.02,Fi.03}. It is based  on the following conjectures:
\begin{enumerate} 
\item 
The nuclear ground state is characterized by strong scalar and vector 
mean fields which have their origin in the in-medium changes of the 
scalar quark condensate (the chiral condensate) and of the quark density. 

\item 
Nuclear binding and saturation arise primarily from chiral (pionic) 
fluctuations (reminiscent of van der Waals forces) in combination with 
Pauli blocking effects and three-nucleon (3N) interactions, 
superimposed on the condensate background fields 
and calculated according to the rules of in-medium chiral perturbation 
theory (ChPT).
\end{enumerate}

The starting point is the description of nuclear matter
based on the chiral effective Lagrangian with pions and 
nucleons~\cite{Kai.01jx,Kai.01ra,Fri.02,Kai.02}, recently 
improved by including
explicit $\Delta (1232)$ degrees of freedom~\cite{Fri.04}.
The relevant ``small'' scales are the Fermi momentum $k_f$,
the pion mass $m_\pi$ and the $\Delta-N$ mass difference $\Delta \equiv
M_\Delta-M_N \simeq 2.1 m_\pi$, all of which are well separated
from the characteristic scale of spontaneous
chiral symmetry breaking, $4\pi f_\pi\simeq 1.16 {\rm ~GeV}$ with the pion
decay constant $f_\pi =92.4 {\rm ~MeV}$. The calculations have been performed 
to three-loop order in the energy density. They incorporate
the one-pion exchange Fock term, iterated one-pion exchange and irreducible
two-pion exchange, including one or two intermediate $\Delta$'s.
The resulting nuclear matter equation of state is
given as an expansion in powers of the Fermi momentum $k_f$.
The expansion coefficients are functions of $k_f/m_\pi$ and $\Delta/m_\pi$,
the dimensionless ratios of the relevant small scales.
Divergent momentum space loop integrals are regularized by introducing
subtraction constants in the spectral representations  of 
these terms~\cite{Fri.04}.
The (few) subtraction constants are the only parameters in this approach.
They equivalently correspond to two- and three-nucleon contact interactions
(and derivatives thereof), encoding short-distance
dynamics not resolved in detail at the 
characteristic momentum scale $k_f \ll 4\pi f_\pi$.
The finite parts of the energy density, written in closed form as functions of
$k_f/ m_\pi$ and $\Delta /m_\pi$, represent long and intermediate range
(chiral) dynamics with input fixed entirely in the $\pi N$ sector.
The low-energy constants (contact terms) are adjusted 
to reproduce basic properties 
of symmetric and asymmetric nuclear matter.

A first version (not yet including explicitly the $\Delta(1232)$) 
of this microscopic approach has been tested in the analysis
of bulk and single-nucleon properties of finite nuclei  \cite{Fi.02,Fi.03}. 
It was shown that chiral (two-pion exchange) fluctuations play
a prominent role in nuclear binding and in the 
saturation mechanism, while additional strong scalar and vector mean fields of
about equal magnitude and
opposite sign, induced by changes of the QCD vacuum in the presence
of baryonic matter, generate the large effective
spin-orbit potential in finite nuclei. A first series of promising results 
for $N\approx Z$ nuclei demonstrated that such an
approach to nuclear dynamics, constrained by
the chiral symmetry breaking pattern and the condensate structure
of low-energy QCD,
has the capability of describing finite nuclei and their 
properties at a quantitative level 
comparable with phenomenological self-consistent mean-field models.

Chiral effective field theories are systematically improved
by introducing explicit $\Delta (1232)$ degrees of freedom. 
Much better isospin properties of nuclear 
matter are found by including chiral $\pi N \Delta$-couplings~\cite{Fri.04}. 
This has an ameliorating influence on the isovector channel 
of the nuclear energy density functional for 
finite nuclei, and much improved results are expected for 
ground-state properties of $N\neq Z$ nuclei. In the present work
the effects of the inclusion of chiral
$\pi N \Delta$-dynamics on the nuclear energy density functional 
will be investigated. Specifically, 
the chiral nuclear matter energy density functional 
will be mapped onto the exchange-correlation energy density 
functional of a covariant point-coupling 
model for finite nuclei, including gradient corrections. 
This model will be employed in the description of ground-state 
properties of a broad range of spherical and deformed nuclei. The results 
will be analyzed in comparison with experimental data on 
binding energies, charge radii, neutron radii and deformation 
parameters for several isotopic chains.

In Section \ref{secII} we construct the nuclear
energy density functional based on the conjectures mentioned previously.
Next, in order to deal with a broad range of finite (medium-heavy and heavy)
nuclei, it is convenient to formulate an equivalent 
covariant point-coupling model 
with density-dependent contact interactions. The mapping of 
the nucleon self-energies in nuclear matter, calculated using 
chiral dynamics, onto 
those of the point coupling model for finite nuclei and the
fine-tuning of the 
remaining parameters is described in Section \ref{secIII}. 
In Section \ref{secIV} the resulting self-consistent equations are solved for 
ground-state properties of a number of spherical and deformed
nuclei. Section \ref{secV} 
summarizes the results of the present investigations and 
ends with an outlook for future studies.

\section{\label{secII}The nuclear energy density functional}

\subsection{Framework and conjectures}

It is useful first to recall some basics of  (non-relativistic) 
Density Functional Theory (DFT)
\cite{HK.64,KS.65,Kohn_Nobel,Dr.Gro} for fermionic many-body systems. 
The starting point of DFT is the Hohenberg-Kohn (HK) variational principle 
for the ground-state energy of interacting fermions:
\begin{equation}
E = \min_\rho \{ F_{\rm HK}[\rho] \} \;.
\end{equation} 
The exact ground-state energy is obtained from the universal functional
$F_{HK}$ depending only on the local density $\rho({\bf r})$.
The Kohn-Sham DFT considers an auxiliary 
system of non-interacting quasi-particles and states that for any interacting 
system, there exists a local single-particle potential (the Kohn-Sham
potential) such that the exact ground-state density of the interacting 
system equals the ground-state density of the auxiliary problem. The 
HK free energy is commonly decomposed into three separate terms:
\begin{equation}
\label{KS}
F_{\rm HK}[\rho] = E_{\rm kin}[\rho] + E_{\rm H}[\rho] + E_{\rm xc}[\rho] \; ,
\end{equation}
where $E_{\rm kin}$ is the kinetic energy of 
the non-interacting N-particle system, $E_{\rm H}$ is a Hartree
energy, and $E_{\rm xc}$  denotes the exchange-correlation 
energy which, by definition, contains everything else.
The practical usefulness of the Kohn-Sham scheme depends entirely
on whether accurate approximations for $E_{\rm xc}$ can be found.
The local ground-state density is constructed using 
so-called auxiliary orbitals,
\begin{equation}
\rho({\bf r}) = \sum_{k=1}^N |\psi_{KS}^k({\bf r})|^2 \; ,
\end{equation}
which are unique functionals of the density $\rho({\bf r})$, i.e.
the KS scheme defines a self-consistency problem.
The set $\{ \psi_{KS}^k\} $ of single-particle wave functions represents
an ``effective'' basis because, in general, these functions do not 
have a directly observable physical 
interpretation\footnote{except for the last 
occupied orbital (see Ref.~\cite{Dr.Gro}).}. 
The corresponding Kohn-Sham single-particle equations, 
which determine $\psi_{KS}^k$, are easy to solve numerically, 
even for large numbers of particles.
The relativistic analogue of the Kohn-Sham scheme, used in 
the present work, starts with a set of Dirac
spinors $\{ \psi_D^k \}$ which replaces the 
non-relativistic wave functions $\{ \psi_{KS}^k\} $.

The conjectures on which our approach to the nuclear 
energy density functional is based, with contact to 
low-energy QCD (see Sec. \ref{secI}), 
can now be adapted as follows:
\begin{enumerate}
\item 
The large scalar and vector mean fields (with opposite signs) 
that have their origin in the in-medium changes of the 
chiral condensate and of the quark density, 
determine the Hartree energy functional $E_{\rm H}[\rho]$.
\item 
The chiral (pionic) fluctuations including one- and
two-pion exchange with single and double 
virtual $\Delta$(1232)-isobar excitations plus Pauli blocking 
effects, determine the 
exchange-correlation energy functional $E_{\rm xc}[\rho]$.
\end{enumerate}
Note that we do not introduce explicit spin orbit terms. 
They emerge naturally from the Lorentz scalar and vector 
self-energies generated by the relativistic density functional, 
as discussed in the following section. Furthermore, contact terms 
from regularization-dependent pieces of two-pion exchange 
processes can be grouped together with the Hartree part $E_{\rm H}$ 
and need not be included in $E_{\rm xc}$. 
However, we prefer to keep them along with the pionic 
fluctuations for convenience.

\subsection{Density-dependent point coupling approach}

The density distribution and the energy of the nuclear ground state 
are obtained from self-consistent solutions of the relativistic 
generalizations 
of the linear single-nucleon Kohn-Sham equations. 
In order to derive those equations it is useful to construct a
point-coupling model~\cite{PC1,PC2,PC3} with density dependent
interaction terms, designed such as to reproduce the detailed
density dependence of the nucleon self-energies resulting from $E_{\rm H}
[\rho] + E_{\rm xc}[\rho]$.

The spin-orbit force in nuclei is well known to be abnormally strong. 
An estimate based on the relation familiar from atoms,
\begin{equation}
\Delta H_{ls} = \frac{{\bf l} \cdot {\bf s}}{2M_N^2 r}
\frac{dU}{dr}
\end{equation}
using the average nuclear single particle potential $U$, 
would give a spin-orbit interaction too weak by about an order of magnitude.
It is also known~\cite{Fri.04} that chiral one- and two-pion exchange dynamics 
alone cannot reproduce the observed ${\bf l} \cdot {\bf s}$ 
interaction. Additional
mechanisms are required, such as the coherent action of 
strong scalar and vector
potentials in relativistic mean-field models. In order
to generate the large empirical spin-orbit splittings, we have two options.
The first one is to stay within a non-relativistic framework
(such as the Skyrme energy density functional) and simply fix the spin-orbit
coupling by an appropriate contact term. The second option 
(the one we adopt here) is to preserve the distinction between scalar and
vector nucleon self-energies in a relativistic description.

In the effective field theory approach to the NN interaction~\cite{Epelbaum_so}
short range central and spin-orbit forces are represented by
contact interactions with low-energy parameters fitted to NN phase shifts. 
It has been demonstrated in~\cite{NK_so} how these low-energy parameters 
translate into an equivalent Skyrme representation 
and - not surprisingly - account for the correct magnitude of the 
spin-orbit force as required already for the fit to triplet p-wave 
NN phase shifts. In our approach the role of these contact terms 
is taken over by the strong Lorentz scalar and vector fields which 
produce effectively the same spin-orbit coupling at the level of the nuclear mean field. 

A successful
framework that meets these requirements for a two component system of protons
and neutrons starts from a relativistic Lagrangian which includes 
isoscalar-scalar (S), isoscalar-vector (V),
isovector-scalar (TS) and isovector-vector (TV)
effective four-fermion interaction vertices with density-dependent 
coupling strengths. The density dependence represents many-body
effects beyond mean field and two-body forces, including 
Pauli blocking effects. In Refs.~\cite{Fi.02,Fi.03} 
we have shown that a very good description of nuclear matter and
finite nuclei from oxygen to calcium can be based on the following 
minimal Lagrangian density:
\begin{equation}
\mathcal{L} = \mathcal{L}_{\rm free} + \mathcal{L}_{\rm int}^{(1)}
  + \mathcal{L}_{\rm int}^{(2)} + \mathcal{L}_{\rm coul},
\label{Lag}
\end{equation}
The four terms read:
\begin{eqnarray}
\label{Lag2}
\mathcal{L}_{\rm free} & = &\bar{\psi}
   (i\gamma_{\mu}\partial^{\mu} -M_N)\psi \; ,\\
\label{Lag3}
\mathcal{L}_{\rm int}^{(1)} & = &
   - \frac{1}{2}~G_{S}(\hat{\rho}) (\bar{\psi}\psi)(\bar{\psi}\psi)
   -\frac{1}{2}~G_{V}(\hat{\rho})(\bar{\psi}\gamma_{\mu}\psi)
   (\bar{\psi}\gamma^{\mu}\psi) \nonumber\\
   & ~ & - \frac{1}{2}~G_{TS}(\hat{\rho})
   (\bar{\psi}\vec{\tau}\psi) \cdot (\bar{\psi} \vec{\tau} \psi)
   - \frac{1}{2}~G_{TV}(\hat{\rho})(\bar{\psi}\vec{\tau}
   \gamma_{\mu}\psi)\cdot (\bar{\psi}\vec{\tau} \gamma^{\mu}\psi) \; ,\\
\label{Lag4}
\mathcal{L}_{\rm int}^{(2)} & = & -\frac{1}{2}~D_{S} \partial_{\nu}(
   \bar{\psi}\psi) \partial^{\nu}(\bar{\psi}\psi) \; , \\
\label{Lag5}
\mathcal{L}_{\rm em} & = &
   eA^{\mu}\bar{\psi}\frac{1+\tau_3}{2}\gamma_{\mu}\psi
   -\frac{1}{4} F_{\mu\nu}F^{\mu\nu} \; ,
\end{eqnarray}
where $\psi$ is the Dirac field of the nucleon with its two isospin 
components (p and n).
Vectors in isospin space are denoted by arrows.
In addition to the free nucleon Lagrangian $\mathcal{L}_{\rm free}$
and the interaction terms contained in
$\mathcal{L}_{\rm int}^{(1)}$, when applied to finite nuclei, the model
must include the coupling $\mathcal{L}_{\rm em}$
of the protons to the electromagnetic field $A^\mu$ with $F_{\mu\nu} =
\partial_\mu A_\nu - \partial_\nu A_\mu$,
and a derivative (surface) term $\mathcal{L}_{\rm int}^{(2)}$. 
One could, of course, construct additional 
derivative terms in $\mathcal{L}_{\rm int}^{(2)}$,
further generalized to include density dependent strength parameters. 
However, as we shall see, there appears to be no need in practical
applications to go beyond the simplest ansatz (\ref{Lag4})
with a constant $D_S$. In fact, present data on nuclear ground states
constrain only this single isoscalar derivative term~\cite{Fur.99}.

The variational principle $\delta\mathcal{L}/\delta \bar{\psi} =0$ applied to
the Lagrangian (\ref{Lag}) leads to the self-consistent single-nucleon Dirac 
equations, the relativistic analogue of the (non-relativistic) 
Kohn-Sham equations. 
The nuclear dynamics produced by chiral (pionic) 
fluctuations in the medium is now encoded in the density dependence of the 
interaction vertices. 

In the framework of
relativistic density functional theory \cite{Dr.Gro,Dr.02,Sp.92,Sc.95},
the density-dependent couplings are functions of the 4-current $j^\mu$:
\begin{equation}
j^\mu = \bar{\psi} \gamma^\mu \psi = \hat{\rho} u^{\mu} \; ,
\end{equation}
where $u^{\mu}$ is the 4-velocity defined as 
$(1-{\bm v}^2)^{-1/2}(1,{\bm v})$. We work in the rest-frame of the 
nuclear system with $\bm v=0$. 

The couplings $G_i(\hat{\rho})$ ($i=S,V,TS,TV$) are decomposed as follows: 
\begin{eqnarray}
\label{coupl}
G_i(\hat{\rho}) &=& G_i^{(0)} + G_i^{(\pi)} (\hat{\rho}) 
\quad ({\rm for}\,\, i=S,V)\nonumber \\
{\rm and} \quad G_i(\hat{\rho}) &=& G_i^{(\pi)} (\hat{\rho}) 
\quad \quad \quad\quad({\rm for}\,\,i=TS,TV)\; ,
\end{eqnarray}
into density-independent parts $G_i^{(0)}$ which arise from strong isoscalar 
scalar and vector background fields, and density-dependent 
parts $G_i^{(\pi)} (\hat{\rho})$ generated by (regularized) one- and two-pion
exchange dynamics. 
As in our previous work, it is assumed that only pionic processes
contribute to the isovector channels.

The relativistic density functional describing the 
ground-state energy of the system
can be re-written as a sum of four distinct terms:
\begin{equation}
\label{dft}
E_0[\hat{\rho}] = 
E_{\rm free} [\hat{\rho}] +
E_{\rm H} [\hat{\rho}] +
E_{\rm coul} [\hat{\rho}] +
E_{\rm \pi} [\hat{\rho}] \; ,
\end{equation}
with
\begin{eqnarray}
\label{e1}
E_{\rm free} [\hat{\rho}] & = & 
\int d^3r~\leftg \bar{\psi} [-i{\bm \gamma} \cdot {\bm \nabla} + M_N ] \psi 
\rightg  \; ,\\
%%%%%%%%%%%%%%%%%%%%%%%%%%%%%%%%%%%%%%%%%%%%%%%%%%%%%%%%%%%%%%
\label{e2}
E_{\rm H} [\hat{\rho}] & = & \frac{1}{2}
\int d^3r~\{ \leftg G_S^{(0)}(\bar{\psi}\psi)^2 \rightg + \leftg
G_V^{(0)} (\bar{\psi} \gamma_\mu \psi)^2 \rightg \} \; , \\ 
%%%%%%%%%%%%%%%%%%%%%%%%%%%%%%%%%%%%%%%%%%%%%%%%%%%%%%%%%%%%%%
\label{e3}
E_{\rm \pi} [\hat{\rho}] & = & \frac{1}{2}
\int d^3r~ \left\{ \leftg G_S^{(\pi)}(\hat{\rho})(\bar{\psi}\psi)^2 \rightg +
\leftg G_V^{(\pi)}(\hat{\rho}) (\bar{\psi} \gamma_\mu \psi)^2 \rightg  \right.
\nonumber \\ 
%%%%%%%%%%%%%%%%%%%%%%%%%%%%%%%%%%%%%%%%%%%%%%%%%%%%%%%%%%%%%%
\label{e4}
& ~ & + \leftg G_{TS}^{(\pi)}(\hat{\rho}) 
(\bar{\psi}\vec{\tau} \psi)^2 \rightg 
+ \leftg G_{TV}^{(\pi)}(\hat{\rho})(\bar{\psi} \gamma_\mu \vec{\tau}
\psi)^2 \rightg  \nonumber \\
%%%%%%%%%%%%%%%%%%%%%%%%%%%%%%%%%%%%%%%%%%%%%%%%%%%%%%%%%%%%%%
\label{e5}
& ~ & - \left. \leftg D_S^{(\pi)} 
[{\bm \nabla}(\bar{\psi}\psi)]^2 \rightg 
%+ \leftg D_V^{\pi} ({\bm \nabla}\bar{\psi} \gamma_\mu \psi)^2 \rightg ] 
\right\} 
\; ,\\
%%%%%%%%%%%%%%%%%%%%%%%%%%%%%%%%%%%%%%%%%%%%%%%%%%%%%%%%%%%%%%
\label{e6}
E_{coul} [\hat{\rho}] & = & \frac{1}{2}
\int d^3r~\leftg A^\mu e\bar{\psi} \frac{1+\tau_3}{2}\gamma_\mu \psi
\rightg \; ~~,
\end{eqnarray}
where $\rightg$ denotes the nuclear ground state.
Here $E_{\rm free}$ is the energy of the free (relativistic)
nucleons including their rest mass. $E_{\rm H}$ is a Hartree-type
contribution representing strong scalar and vector mean fields,
later to be connected with the leading terms of the corresponding nucleon
self-energies deduced from in-medium QCD sum rules. Furthermore,
$E_{\rm \pi}$ is the part of the energy generated by chiral $\pi N
\Delta$-dynamics, including a derivative (surface) term, with all pieces
explicitly derived in \cite{Fri.04}.

Note that Eqs.~(\ref{dft}-\ref{e6}) are arranged in close 
correspondence with a
relativistic extension of the Hohenberg-Kohn energy functional (\ref{KS}), 
identifying the Hartree term $E_{\rm H}$ with Eq.~(\ref{e2}) plus the
Coulomb energy (\ref{e6}), and the exchange correlation part with 
$E_{\rm \pi}$ of Eq.~(\ref{e4}).
Minimization of the ground-state energy, represented in terms of a set of
auxiliary Dirac spinors ${\psi_k}$, leads to the relativistic analogue of the
Kohn-Sham equations. These single-nucleon Dirac equations are solved
self-consistently in the ``no-sea''
approximation which omits the explicit contribution 
of the negative-energy solutions of the relativistic 
equations to the densities and currents. This means that 
vacuum polarization effects are not taken into account 
explicitly. They are included in the adjustable parameters of the theory
(for an exhaustive discussion 
see Refs.~\cite{Dr.02,QHD,Ring_lectures,Fur.03}). 
This no-sea approximation is well founded within the framework
of an effective field theory. Nucleon-antinucleon
fluctuations occur at high-energy (short-distance)
scales, not resolved at the long wavelength characteristic
of the nuclear Fermi momentum. Such fluctuations are implicitly absorbed,
together with other short-range dynamics, in contact terms of 
the effective Lagrangian.

The expressions for the isoscalar and isovector four-currents and 
scalar densities read:
\begin{eqnarray}
\label{den1}
j_\mu & = \leftg \bar{\psi} \gamma_\mu \psi \rightg =
& \sum_{k=1}^N \bar{\psi}_k \gamma_\mu \psi_k \; ,\\
\label{den2}
\vec{j}_\mu & =  
\leftg \bar{\psi} \gamma_\mu \vec{\tau} \psi \rightg =
& \sum_{k=1}^N \bar{\psi}_k \gamma_\mu \vec{\tau} \psi_k \; ,\\
\label{den3}
\rho_S & = \leftg \bar{\psi} \psi \rightg = 
& \sum_{k=1}^N \bar{\psi}_k \psi_k \; ,\\
\label{den4}
\vec{\rho}_S & = \leftg \bar{\psi} \vec{\tau} \psi \rightg =
& \sum_{k=1}^N \bar{\psi}_k \vec{\tau} \psi_k \; ,
\end{eqnarray}
where $\psi_k$ are Dirac spinors and
the sum runs over occupied positive-energy single-nucleon states.

In this work we only consider systems with time-reversal
symmetry in the ground-state, i.e. even-even nuclei. 
The space components of all 
currents vanish (${\bf j}=0$), and because of
charge conservation only the third component of
isospin ($\tau_3 = -1$ for neutrons and $ \tau_3 = +1 $ for protons)
contributes. The relevant combinations of densities are:
\begin{eqnarray}
   \rho & = & \leftg \bar{\psi}\gamma^{0} \psi \rightg
    = \rho^p + \rho^n \; , \\
   \rho_{3} & = & \leftg \bar{\psi} \tau_3 \gamma^{0} 
   \psi \rightg
   = \rho^p - \rho^n  \; ,\\
   \rho_S & = & \leftg \bar{\psi} \psi \rightg 
    = \rho_S^p + \rho_S^n \; , \\
   \rho_{S3} & = & \leftg \bar{\psi} \tau_3 \psi \rightg 
    = \rho_S^p - \rho_S^n \; .
\end{eqnarray}
The charge density is 
\begin{eqnarray}
\rho_{ch} = \leftg \bar{\psi}\gamma^{0}{1+\tau_3\over 2} 
\psi \rightg = \rho^p ~. 
\nonumber
\end{eqnarray}
The single-nucleon Dirac equation is found
by minimization with respect to $\bar{\psi}_k$:
\begin{eqnarray}
\label{mean-field}
[-i{\bm \gamma}\cdot {\bm \nabla} + M_N
+ \gamma_0 \Sigma_V &+& \gamma_0 \tau_3 \Sigma_{TV} \nonumber\\
&+&\gamma_0 \Sigma_R + \Sigma_S + \tau_3 \Sigma_{TS}
] \psi_k = \epsilon_k \psi_k \; ,
\end{eqnarray}
with the self-energies:
\begin{eqnarray}
\label{self1}
\Sigma_V & = & [G_V^{(0)} + G_V^{(\pi)}(\rho)] \rho 
+ eA^0\frac{1+\tau_3}{2}\; ,\\
\label{self2}
\Sigma_{TV} & = & G_{TV}^{(\pi)}(\rho)\, \rho_{3} \; ,\\
\label{self3}
\Sigma_{S}   & = & [G_S^{(0)} + G_S^{(\pi)}(\rho)] \rho_S + D_S^{(\pi)}
{\bm \nabla}^2 \rho_S \; ,\\
\label{self4}
\Sigma_{TS} & = & G_{TS}^{(\pi)}(\rho)\, \rho_{S3} \; ,\\
\label{self5}
\Sigma_R & = & \frac{1}{2} \left\{ \frac{\partial G_V^{(\pi)}(\rho)
}{\partial \rho}
\rho^2 +   \frac{\partial G_S^{(\pi)}(\rho)
}{\partial \rho} \rho_S^2\, + \right. \nonumber \\
 & ~ & \left. \hspace{3cm} \frac{\partial G_{TV}^{(\pi)}(\rho)
}{\partial \rho}
\rho_{3}^2 + \frac{\partial G_{TS}^{(\pi)}(\rho)
}{\partial \rho}
\rho_{S3}^2 \right\} \; ,
\end{eqnarray}
where $A^0 ({\bf r}) = \frac{e}{4\pi} \int d^3r'~\frac{\rho_{ch} 
({\bf r'})}{|{\bf r} - {\bf r'}|}$ is the Coulomb potential.
In addition to the usual contributions from the time components of the 
vector self-energies and the scalar potentials, we must also 
include the ``rearrangement'' 
terms, $\Sigma_R$, arising from the variation of the vertex 
functionals with respect to the nucleon fields
in the density operator $\hat{\rho}$. 
For a Lagrangian with density dependent couplings, the inclusion
of the rearrangement self-energies
is essential in order to guarantee energy-momentum conservation and 
thermodynamical consistency \cite{FLW.95,Len.04,TW.99}. It is also
required by the Hugenholtz-Van Hove 
theorem~\cite{Hugen}. Using the single-nucleon Dirac equation and
performing an integration by parts, the ground-state energy of a 
nucleus with A nucleons reads:
\begin{eqnarray}
  E_0 = 
   \sum\limits_{k=1}^{A} \epsilon_k & ~ & 
   - \frac{1}{2} \int d^3r \left\{
   ~[G_S^{(0)} + G_S^{(\pi)}(\rho)] \,\rho_S^2 + 
    G_{TS}^{(\pi)}(\rho) \,\rho_{S3}^2 + \right. \nonumber\\
   & ~ &  [G_V^{(0)} + G_V^{(\pi)}(\rho)]\, \rho^2 + 
    G_{TV}^{(\pi)}(\rho)\, \rho_{3}^2 + \nonumber\\
 & ~ & 
   \frac{\partial G_S^{(\pi)}(\rho)}{\partial \rho}\,\rho_S^2\, \rho +
   \frac{\partial G_{TS}^{(\pi)}(\rho)}{\partial \rho} \,\rho^2_{S3}\, \rho +
   \nonumber\\ 
 & ~ &
   \frac{\partial G_V^{(\pi)}(\rho)}{\partial \rho}\, \rho^3 +
   \frac{\partial G_{TV}^{(\pi)}(\rho)}{\partial \rho} \,\rho^2_{3}\, 
   \rho + \nonumber\\
 & ~ & \left. e\,\rho_{ch}\, A^0 + D_S^{(\pi)} \rho_S \nabla^2 \rho_S 
   \right\} \; ,  
\end{eqnarray}
where $\epsilon_k$ denotes the single-nucleon energies.

%%%%%%%%%%%%%%%%%%%%%%%%%%%%%%%%%%%%%%%%%%%%%%%%%%%%%%%%%%%%%%%%%%%%%%%%%%%%%

\section{\label{secIII}Linking the energy functional to the low-energy sector
of QCD.}
 
\subsection{\label{secIIIa}Scalar and vector fields: guidance from in-medium 
QCD sum rules.} 

The QCD ground state (or ``vacuum") is characterized by strong condensates of 
quark-antiquark pairs 
and gluons, an entirely non-perturbative phenomenon. The quark condensate
$\langle \bar{q}q \rangle$, i.e. the ground state expectation value of the
scalar quark density, plays a particularly important role 
as an order parameter
of spontaneously broken chiral symmetry. At a renormalization scale
of about 1 GeV (with up and down quark masses $m_u + m_d \simeq 12$ MeV) 
the value of the chiral vacuum condensate \cite{Pic} is 
$\langle \bar{q} q \rangle_0 \simeq - (240~
{\rm MeV})^3 \simeq - 1.8~{\rm fm}^{-3}$.
Hadrons, as well as nuclei, are excitations built on this condensed ground
state. Changes of the condensate structure of the QCD vacuum in the presence
of baryonic matter are a source of strong fields 
experienced by the nucleons in the bulk of nuclei. 
 
In-medium QCD sum rules relate the leading changes of the scalar quark 
condensate and of the quark density at finite baryon density, 
with the scalar and vector self-energies
of a nucleon in the nuclear medium. 
To first order in the scalar and baryon densities, these 
self-energies can be expressed as 
follows~\cite{Fi.03,Co.91,Fur.92,Co.94,DL.90,DL.91}:
\begin{eqnarray}
\label{back1}
   \Sigma_S^{(0)} & = & 
   - \frac{\sigma_N M_N}{m_\pi^2 f_\pi^2} \rho_S \; , \\
\label{back2}
   \Sigma_V^{(0)} & = & \frac{4 (m_u + m_d)M_N}{m_\pi^2f_\pi^2} \rho 
\; ,
\end{eqnarray}
where $\sigma_N = \langle N| m_q \,\bar{q} q |N \rangle$ is
the nucleon sigma term ($\simeq 50$ MeV), $m_\pi$ is the pion mass (138 MeV),
and $f_\pi = 92.4$ MeV is the pion decay constant.
The resulting $\Sigma_S^{(0)}$ and $\Sigma_V^{(0)}$ are 
individually of the order of
300 -- 400 MeV in magnitude. Their ratio 
\begin{equation}
\label{ratio}
   \frac{\Sigma_S^{(0)}}{\Sigma_V^{(0)}} = - \frac{\sigma_N}{4(m_u +m_d)}
   \frac{\rho_S}{\rho}
\end{equation}
is close to $-1$, suggesting a large cancellation of scalar and
vector potentials in the single-nucleon Dirac equation, 
a feature characteristic 
of relativistic mean-field phenomenology. Of course the constraints
implied by Eqs. (\ref{back1})-(\ref{back2}), and 
by the ratio Eq. (\ref{ratio})
are not very accurate on a quantitative level. Beyond
the leading-order Ioffe formula~\cite{{Iof}} on which Eq. (\ref{back1}) 
is based, one finds corrections from condensates of higher dimension 
which are not well under control. Given in addition 
the uncertainties in the values of
$\sigma_N$ and $m_u+m_d$, the estimated error for the ratio 
$\Sigma_S^{(0)}/\Sigma_V^{(0)}\simeq -1$ is about 20\%.
It is important to note, however, that the explicitly calculated 
two-pion exchange fluctuations, used in this work, account for 
a large part of the effects from higher 
dimensional condensates (such as the four-quark
condensate). Restricting $\Sigma_{S,V}^{(0)}$ to the leading order terms
(\ref{back1}-\ref{back2}) does therefore make sense.

Comparing Eqs. (\ref{self1}) and (\ref{self3}) for the isoscalar vector and 
scalar potentials of the single-nucleon Dirac equations, with the Eqs.
(\ref{back1}) and (\ref{back2}) for the condensate background self-energies,
respectively, the following estimates hold for the 
couplings of the nucleon to the background fields (the Hartree terms in the 
energy functional):
\begin{equation}
G_S^{(0)} = - \frac{\sigma_N M_N}{m_\pi^2 f_\pi^2} \; ,\\
\end{equation}
and
\begin{equation}
G_V^{(0)} = \frac{4(m_u + m_d)M_N}{m_\pi^2 f_\pi^2} \; .
\end{equation}
Inserting the values of the nucleon and pion masses and the pion
decay constant,
\begin{eqnarray}
\label{estim1}
G_S^{(0)} & \simeq & - 11~{\rm fm}^{2}~\left[ \frac{\sigma_N}{50~{\rm MeV}} 
\right] \;, \\
\label{estim2}
G_V^{(0)} & \simeq & 11~{\rm fm}^{2}~\left[ \frac{4(m_u + m_d)}{50~{\rm MeV}} 
\right] \;,
\end{eqnarray}
which implies $G_S^{(0)} \simeq - G_V^{(0)} \simeq -10.6~{\rm fm}^2$ 
for typical values $\sigma_N \simeq 48~{\rm MeV}$ and $m_u + m_d 
\simeq 12~{\rm MeV}$ (at a renormalization scale of about 1 GeV). 
This estimate is useful for first orientation. In the actual
applications to finite nuclei, $G_{S,V}^{(0)}$ will have to be fine-tuned.
We mention here in advance that, as a remarkable and non-trivial result 
of this fine-tuning, deviations from the ``canonical'' values
(\ref{estim1}-\ref{estim2}) will turn out to be surprisingly small.

%%%%%%%%%%%%%%%%%%%%%%%%%%%%%%%%%%%%%%%%%%%%%%%%%%%%%%%%%%%%%%%%%%%%%%%%%%%%%%

\subsection{\label{secIIIb}The exchange-correlation term: In-medium chiral 
perturbation theory.}

The many-body effects represented by the exchange-correlation 
density functional are approximated by chiral 
$\pi N \Delta$-dynamics, including Pauli blocking effects. 
Our framework is in-medium chiral perturbation theory with 
inclusion of $\Delta$-isobars, computed to three-loop 
order in the energy density and described in detail 
in Refs.~\cite{Kai.01jx,Kai.01ra,Fri.02,Kai.02,Fri.04}.

Regularization dependent short-range 
contributions from pion loops are encoded in a few
parameters representing counter terms,
or equivalently, subtraction constants in spectral 
representations of the respective momentum space amplitudes. 
At the level of the effective Lagrangian, they appear as 
contact interactions or derivatives thereof. The counter-term 
contributions to the energy per particle, ${\bar E} = E/A - M_N$, 
in symmetric nuclear matter are written: 
\begin{equation}
\Delta\bar{E}(k_f)^{(ct)} = b_{3} \frac{k_f^3}{\Lambda^2} + b_{5} 
\frac{k_f^5}{\Lambda^4} + b_{6} 
\frac{k_f^6}{\Lambda^5}\; .
\label{de}
\end{equation}
At this stage the calculation for homogeneous nuclear matter 
excludes surface terms which are discussed in the next section. 
The constants $b_{i}$ are defined dimensionless, along with a 
convenient scale $\Lambda$ to be chosen larger than the 
"small" scales ($k_f, m_\pi$ and $\Delta$) but smaller than 
the "chiral gap", $4\pi f_\pi$. Our choice is 
$\Lambda = 2\pi f_\pi \simeq 0.58$ GeV.

For asymmetric nuclear matter, the energy per particle 
is written $\bar{E}_{as} = \bar{E} + \delta^2 S_2$ with
 the asymmetry energy $S_2$ and $\delta = (\rho^n - \rho^p)/\rho$. 
The counter-term contributions to the asymmetry energy has 
an expansion analogous to (\ref{de}),
\begin{equation}
\Delta S_2(k_f)^{(ct)} = a_{3} \frac{k_f^3}{\Lambda^2} + a_{5} 
\frac{k_f^5}{\Lambda^4} + a_{6} 
\frac{k_f^6}{\Lambda^5}\; ,
\label{das}
\end{equation}
with dimensionless constants $a_i$. The parameters $b_i$ 
and $a_i$ subsume unresolved 
short-distance NN-dynamics in the nuclear medium. Note that 
no regularization is required at order $k_f^4$ where the 
corresponding loop integrals are all finite. There are no 
contact interactions that generate a $k_f^4$ term at the 
Hartree-Fock level; gradient terms start minimally as $\nabla^2$.
At order $k_f^5$ the constants $b_5$ and $a_5$ 
represent short-distance effects partly arising from momentum 
dependent interactions. In actual practice it turns out that 
these constants can be set to zero \cite{Fri.04}, the primary 
momentum dependent forces being already well described by the 
finite parts of in-medium two-pion exchange pieces. Additional 
short-range three-body contributions and effects of higher 
loops which feed back into $k_f^6$ terms are parameterized 
by the constants $b_6$ and $a_6$, respectively. 

In Ref. \cite{Fri.04} the counter terms $b_i$ and $a_i$ have 
been adjusted to reproduce nuclear and neutron matter properties. 
The resulting equation of state of isospin-symmetric matter is in good 
agreement with recent microscopic calculations \cite{Pa.}, 
though with a somewhat too high incompressibility ($K_0 = 304$ MeV).
The calculated real part of the nucleon single-particle 
potential, $U(p,k_f)$, 
is very close to the result of relativistic Dirac-Brueckner 
calculations (see Ref.~\cite{Li.92}). Isospin properties of 
nuclear matter and the energy per particle in neutron matter 
are significantly improved by incorporating explicit 
$\pi N \Delta$-dynamics, in comparison with earlier calculations 
(see Ref.~\cite{Kai.01jx})
which included only nucleons in two-pion exchange processes.

While gross properties of infinite nuclear matter are useful for 
orientation, the large amount of nuclear observables studied in 
the present work provides a far more accurate data base that 
permits a better adjustment of the constants $b_i$ and $a_i$. 
We refer to the more detailed discussion in Section 3.3 but 
point here already to the interesting result that the best fit 
values for $b_3$ and $a_3$ turn out to be within only a few 
percent of those determined in the ChPT calculation of nuclear 
and neutron matter, while $b_5 = a_5 = 0$ can still be 
maintained. The only major difference is in the short-distance 
three-body term proportional to $b_6$ for which the fit to the 
broad range of nuclear data requires stronger attraction. 
Throughout the procedure, the input values for the chiral 
pion-nucleon and $\pi N \Delta$ couplings are strictly kept 
fixed by pion-nucleon scattering observables in vacuum. 
These couplings determine the finite parts of intermediate-range 
one- and two- pion exchange contributions to the energy density 
as predicted by in-medium ChPT.

In the simplest DFT approach, the exchange-correlation 
energy for a finite system is determined 
in the local density approximation (LDA)
from the exchange-correlation functional of the corresponding 
infinite homogeneous system,
replacing the constant density $\rho$ by the
local density $\rho ({\bf r})$ of the actual inhomogeneous system.
In our case the exchange-correlation terms of the nuclear density
functional are determined within LDA by equating the 
corresponding self-energies in the single-nucleon Dirac equation 
(\ref{mean-field}), with those arising from the in-medium 
chiral perturbation theory calculation of 
$\pi N \Delta$-dynamics in homogeneous isospin symmetric and
asymmetric nuclear matter. Steps beyond the LDA will be taken by adding
surface terms involving derivatives of the density. 

The density-dependent couplings 
$G^{(\pi)}_i$ are expressed as polynomials in fractional powers of
the baryon density: 
\begin{equation}
G^{(\pi)}_i (\rho) = c_{i1} + c_{i2} \rho^{\frac{1}{3}} + c_{i3} 
\rho^{\frac{2}{3}} + c_{i4} \rho \ldots ~(i= S,V,TS,TV) \; .
\end{equation}
The detailed derivation of the constants $c_{ij}$ 
is presented in the Appendix.

The coefficient $D_{S}^{(\pi)}$ of the derivative term in the 
equivalent point-coupling model (Eqs. (\ref{self1}) and (\ref{self3}))
can be determined from ChPT calculations for inhomogeneous nuclear matter.
The density-matrix expansion method \cite{Ne.Va} is used to
derive the in-medium insertion in the nucleon propagator, as shown
in Refs. \cite{Kai.01ra,Fri.04}.
The isoscalar nuclear energy density emerging 
from chiral pion-nucleon dynamics has the form:
\begin{eqnarray} 
\label{Skyrme}
{\mathcal E}(\rho, {\bm \nabla }\rho)& = & \rho\,\bar E(k_f) 
+ ({\bm \nabla }\rho)^2 \, F_\nabla(k_f) + \ldots \; . 
\end{eqnarray} 
The coefficients have been compared with the corresponding phenomenological 
parameters of various Skyrme type energy density functionals. In particular,
the gradient term $ ({\bm \nabla }\rho)^2 F_\nabla(k_f)$ plays an important 
role in shaping   
the nuclear surface~\cite{Va.Br}. While the applicability of 
the density-matrix expansion is questionable at very low densities, 
an important result of Refs. \cite{Kai.01ra,Fri.04} is that around 
nuclear matter saturation density, the
parameter-free ChPT prediction for $F_\nabla(k_{f0})$ is in very good 
agreement with the empirical values used in standard parameterizations
of the Skyrme density functional \cite{BHR.03}. We can approximate 
$F_\nabla$ by a constant in the relevant region of densities. The 
following relationship between $F_\nabla$ and the derivative term 
of the point-coupling model (see Eqs.~(\ref{Lag4},\ref{e5})) is then 
deduced by a straightforward comparison \cite{SavToki}:
\begin{equation}
- 2 {F_\nabla} = D_S^{(\pi)} \; .
\label{ds} 
\end{equation} 
With the average value of $F_\nabla(k_f)$ in the region 
$0.1$ fm$^{-3} \le \rho \le 0.2$ fm$^{-3}$ \cite{Fri.04}, we
estimate $D_S^{(\pi)} = - 0.7~{\rm fm}^4$. 

The inclusion of derivative terms in the model Lagrangian and the 
determination of its strength parameters from ChPT 
calculations in inhomogeneous matter actually goes beyond the 
local density approximation. The term, Eq.~(\ref{Lag4}), 
with the strength parameter given by Eq.~(\ref{ds}), represents a
second-order gradient correction to the LDA, i.e. the next-to-leading term 
in the gradient expansion of the exchange-correlation energy calculated 
by in-medium chiral perturbation theory. 

%%%%%%%%%%%%%%%%%%%%%%%%%%%%%%%%%%%%%%%%%%%%%%%%%%%%%%%%%%%%%%%%%%%%%%%%%%%%
\begin{table}
\begin{center}
\caption{{\it Empirical} nuclear matter properties (energy per particle, 
saturation density, incompressibility, Dirac effective mass, 
symmetry energy) employed in the adjustment of parameters. 
For each ``empirical'' quantity 
we include the range of values within which the 
quantity was allowed to vary during the fit.}
\bigskip
\begin{tabular}{|c|c|c|c|c|c|}
\hline
   ~ & $E/A$ (MeV) & $\rho_{sat}$ ({\rm fm$^{-3}$}) & $K_0$ (MeV) &
     $M^*/M_N$  & $S_2$ (MeV) \\
\hline
Fit   & -16. ~~($\pm 1.$) & 0.15~~ ($\pm 0.01$) & 250.~~ ($\pm 25.$) & 
0.6~~ ($\pm 0.5$) & 33. ~~($\pm 3.$) \\
\hline
\end{tabular}
\label{tab1}
\end{center}
\end{table}
%%%%%%%%%%%%%%%%%%%%%%%%%%%%%%%%%%%%%%%%%%%%%%%%%%%%%%%%%%%%%%%%%%%%%%%%%%%%

%%%%%%%%%%%%%%%%%%%%%%%%%%%%%%%%%%%%%%%%%%%%%%%%%%%%%%%%%%%%%%%%%%%%%%%%%%%%
\begin{table}
\begin{center}
\caption{Empirical properties of finite nuclei ~\cite{Audi2003,NMG.94} 
employed in the fitting procedure:
binding energies ($E_b$), charge radii ($r_{ch}$) and 
differences between r.m.s. neutron and proton radii ($r_n - r_p$). The
relative weights used in fitting these quantities 
are $0.3 \%$, $0.2 \%$ and $10 \%$ respectively.
}
\bigskip
\begin{tabular}{|c|c|c|c|}
\hline
~             & $E_b$ (MeV)   & $r_{ch}$ (fm) & $r_n-r_p$ (fm) \\
\hline
~$^{16}$O     & -127.62  &  2.73        & ---   \\
~$^{40}$Ca    & -342.05  &  3.49        & ---   \\
~$^{48}$Ca    & -415.99  &  3.48        & ---   \\
~$^{72}$Ni    & -613.17  &  ---         & ---   \\
~$^{90}$Zr    & -783.89  &  4.27        & ---   \\
~$^{116}$Sn   & -988.68  &  4.63        & 0.12 \\
~$^{124}$Sn   & -1049.96 &  4.67        & 0.19 \\
~$^{132}$Sn   & -1102.92 &  ---         & ---   \\
~$^{204}$Pb   & -1607.52 &  5.49        & ---   \\
~$^{208}$Pb   & -1636.45 &  5.51        & 0.20 \\
~$^{214}$Pb   & -1663.30 &  5.56        & ---   \\
~$^{210}$Po   & -1645.23 &  ---         & ---   \\
\hline
\end{tabular}
\label{tab2}
\end{center}
\end{table}
%%%%%%%%%%%%%%%%%%%%%%%%%%%%%%%%%%%%%%%%%%%%%%%%%%%%%%%%%%%%%%%%%%%%%%%%%%%%%

\subsection{\label{secIIIc}Adjusting parameters}

The parameters of the point-coupling model are fixed 
simultaneously to properties of nuclear matter (see
Table~\ref{tab1}), and to binding energies, charge radii and
differences between neutron and proton radii of spherical nuclei
(see Table~\ref{tab2}). The inclusion of neutron-rich nuclei, in
particular, provides strict constraints for the parameters 
in the isovector channel. In both tables the empirical values and 
data are listed together with the relative errors used in the 
fitting procedure. 

Corrections have been applied for center-of-mass motion and pairing. 
From the solution of the self-consistent equations we 
subtract the microscopically
calculated center-of-mass correction from the total 
binding energy, following \cite{Rei}:
\begin{equation}
\label{cms}
\Delta E_{cm} = - \frac{\langle P_{cm}^2 \rangle}{2AM_N} \;,
\end{equation}
where $P_{cm}^2$ is the total squared momentum 
of a nucleus with $A$ nucleons.
For the open shell nuclei in Table~\ref{tab2},
pairing correlations are treated
in the BCS approximation with empirical gaps \cite{Mo.Ni}. 

We have adjusted a minimal set of parameters starting from the 
estimates Eqs. (\ref{estim1}) and (\ref{estim2}) for the couplings 
of the condensate background fields (Hartree term), and 
the counter term constants $b_{3}$, $a_{3}$, 
$b_{5}$, $a_{5}$ and $b_6$, $a_6$ guided by Ref.~\cite{Fri.04} 
for the self-energies arising from chiral
$\pi N \Delta$-dynamics (exchange-correlation term). 
In particular we set $b_{5}=a_{5}=0$ as in Ref.~\cite{Fri.04}.
The total number of adjustable parameters is seven. However, five of those
will turn out to be remarkably close to preceding expectations 
or predictions. Only the two
constants associated with three-body correlations are free of links to other
constraints.

The resulting optimal parameter set (FKVW) is shown in Table~\ref{tab3} 
in comparison with estimates and predictions: 
a) for the Hartree term  from in-medium QCD sum rules; b) for
the exchange-correlation functional from the in-medium ChPT calculations of 
$\pi N \Delta$-dynamics in nuclear matter. 

\begin{table}
\begin{center}
\caption{Best-fit parameters of the density-dependent 
point coupling model (Fit FKVW).
Upper panel: coupling strengths of background (Hartree) terms 
in comparison with estimates from in-medium QCD sum rules 
(\ref{estim1}, \ref{estim2}) (QCDSR). Lower panel: constants 
of counter terms from short-distance regularization (reference 
scale $\Lambda = 2\pi f_\pi \simeq 0.58$ GeV) and surface 
(derivative) term $D_S^{(\pi)}$ in comparison with results 
from in-medium ChPT at three-loop order including $\Delta(1232)$ 
\cite{Fri.04}. The choice $b_5 = a_5 = 0$ from Ref.~\cite{Fri.04} 
has been kept, and these parameters are not included in the 
fitting procedure.   
}
\bigskip
\begin{tabular}{|c|}
\hline
Background (Hartree) terms\\
\hline
\begin{tabular}{c|c|c}
~ & Fit FKVW & QCDSR\\
$G_S^{(0)}$ [fm$^2$]& - 11.5  & -10.6 \\
$G_V^{(0)}$ [fm$^2$]&   11.0  &  10.6 \\
\end{tabular}\\
\hline
Counter terms / Pionic fluctuations\\
\hline
\begin{tabular}{c|c|c}
~ & Fit FKVW & ChPT  \cite{Fri.04}\\
$b_3$               &  -2.93 & -3.05  \\
$a_3$               &   2.20 &  2.16  \\
$b_{5}^{*}$         &   0    &  0     \\
$a_{5}^{*}$         &   0    &  0     \\
$b_6$               &  -5.68 & -2.83  \\
$a_6$               &  -0.13 &  2.83  \\
$D_S^{(\pi)}$ [fm$^4$]     &  -0.76  & -0.7  \\
\end{tabular}\\
\hline
$~^{*}$: inputs, see Ref.~\cite{Fri.04}\\
\hline
\end{tabular}
\label{tab3}
\end{center}
\end{table}
%%%%%%%%%%%%%%%%%%%%%%%%%%%%%%%%%%%%%%%%%%%%%%%%%%%%%%%%%%%%%%%%%%%%%%%%%%%% 

Evidently the {\it best fit} parameters are remarkably close 
to the anticipated QCD sum rule and ChPT values,
with the exception of the $k_f^6$ terms ($b_6$ and $a_6$) for which 
the fit to nuclear data systematically requires an attractive shift 
as compared to the ChPT calculation \cite{Fri.04}.

The QCD sum rule predictions for the 
condensate background scalar and vector self-energies have 
an accuracy of about 20 \% as mentioned previously. 
Surprisingly, however, the leading-order sum rule 
estimates, Eqs. (\ref{estim1}) and (\ref{estim2}), 
turn out to be realized remarkably well when 
the parameters of the scalar and vector Hartree fields 
are varied freely and optimized in confrontation
with a large number of high-precision nuclear data. 
There appears to be an a posteriori consistency with 
the implicit assumption made in the in-medium ChPT 
calculations based on $\pi N \Delta$-dynamics: 
namely that the individually large scalar and 
vector nucleon self-energies  
$\Sigma_S^{(0)}$ and $\Sigma_V^{(0)}$ cancel 
in their contribution to the energy density
around saturation. Of course, the $b_3$ counter 
term at order $k_f^3$ from regularized two-pion 
exchange can be translated into an equivalent Hartree 
term as well, such that the total balance of
scalar and vector self-energy contributions 
linear in density is attractive and reminiscent
of QHD and the Walecka model \cite{QHD}. The 
repulsive term of order $k_f^4$ (or $\rho^{4/3}$), 
on the other hand, results
model-independently from Pauli blocking 
effects on chiral two-pion exchange within 
in-medium ChPT and has no counterpart in 
relativistic mean field models.

It is instructive to examine the hierarchy of 
terms in the $k_f$-expansion of the nucleon isoscalar 
single particle potential,
\begin{equation}
U(k_f) = g_3 \frac{k_f^3}{\Lambda^2} + g_4 \frac{k_f^4}{\Lambda^3} +
g_5 \frac{k_f^5}{\Lambda^4} + 
g_6 \frac{k_f^6}{\Lambda^5} + \mathcal{O}\big(\frac{k_f^7}{\Lambda^6}\big)~~, 
\label{sppot}
\end{equation}
given in detail in the Appendix. The reference 
scale is chosen again as $\Lambda = 2\pi f_\pi \simeq 0.58$ GeV. 
The resulting coupling strengths $g_i$ 
are listed in Table \ref{tab4}. They are all of 
"natural" size when looked at from the point of view 
of an effective field theory. Background fields 
can be ignored here because their contributions 
to $U$ cancel almost completely, i.e.  
$U^{(0)} = G^{(0)}_S \rho_S + G^{(0)}_V \rho \simeq 0$. 

%%%%%%%%%%%%%%%%%%%%%%%%%%%%%%%%%%%%%%%%%%%%%%%%%

\begin{table}
\begin{center}
\caption{Comparison between fine-tuned coefficients $g_i$ 
(Fit FKVW) in the expansion
(\ref{sppot}) of the isoscalar single-particle potential,
and ChPT values of $g_i$ deduced from calculations performed in
Ref.~\cite{Fri.04}. The constants $b_i$ used in this compilation
 are those given in Table \ref{tab3} (see Appendix for further details).
}
\bigskip
\begin{tabular}{|c|c|c|}
\hline
 & Fit FKVW & ChPT~\cite{Fri.04} \\
\hline
$g_3$  & -2.84 & -3.10 \\
\hline
$g_4$  & 2.67  &  2.67 \\
\hline
$g_5$  & 2.17  &  2.17 \\ 
\hline
$g_6$  & -3.35 &  5.21 \\
\hline
\end{tabular}
\label{tab4}
\end{center}
\end{table}
%%%%%%%%%%%%%%%%%%%%%%%%%%%%%%%%%%%%%%%%%%%%%%%%%%%%%%%%%%%%%%%%%%%%%%%%%%%%%

\begin{figure}
\includegraphics[scale=0.55,angle=0]{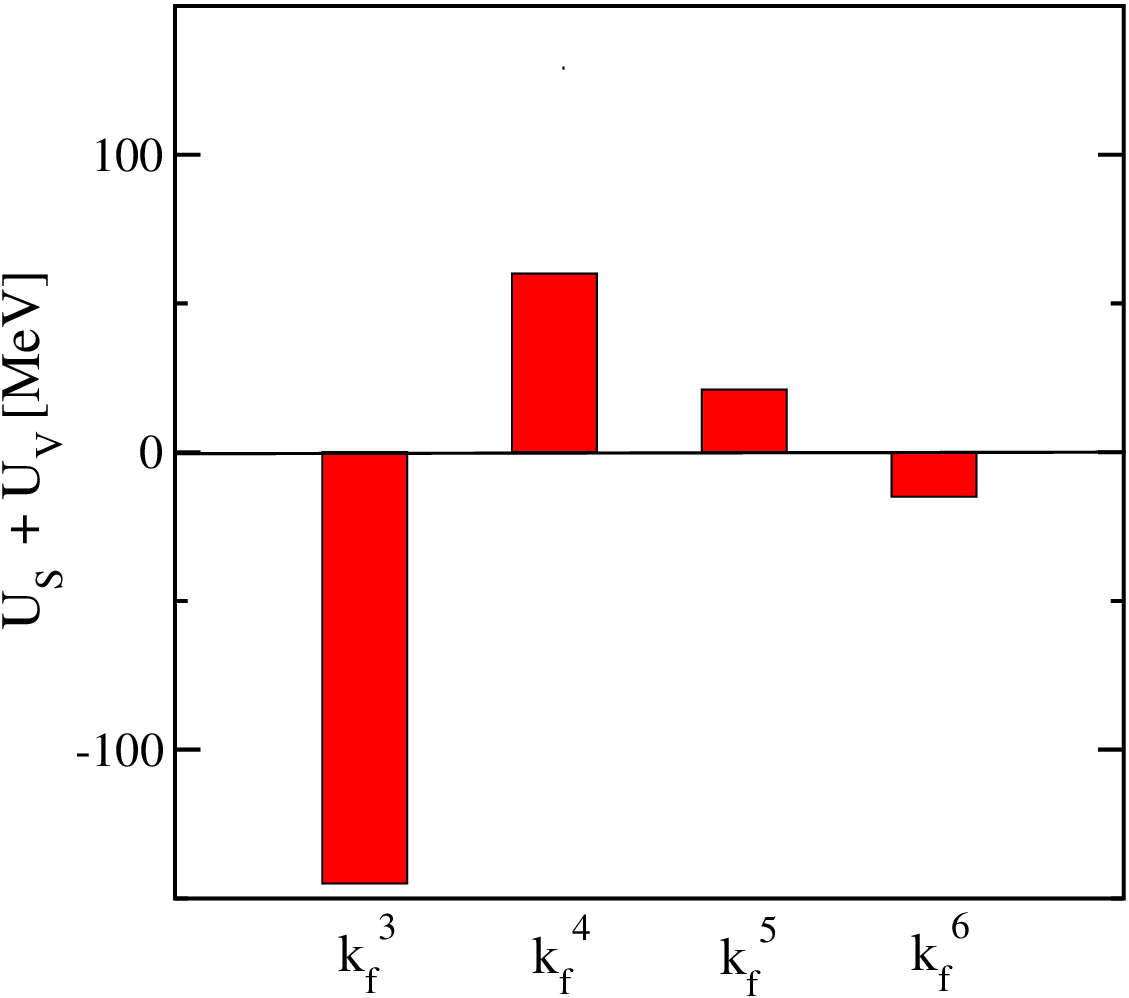}
\vspace{0.5 cm}
\caption{\label{figU}
{Coefficients $U_n$ in the expansion (\ref{sppot2}) of the nuclear single
particle potential in fractional powers $n/3$ of the density $\rho({\bf r})$ 
for the case of a density $\rho \simeq \rho_{sat}$.}}
\end{figure}

An alternative way of displaying the convergence properties 
of the $k_f$ expansion is by translating $U$ of Eq. (\ref{sppot}) 
into the following form of a (local) potential:
\begin{equation}
U({\bf r}) = U_3 \frac{\rho({\bf r})}{\rho(0)} + U_4 
\left(\frac{\rho({\bf r})}{\rho(0)}\right)^{4/3} +
U_5  \left(\frac{\rho({\bf r})}{\rho(0)}\right)^{5/3} + 
U_6 \left(\frac{\rho({\bf r})}{\rho(0)}\right)^2~~, 
\label{sppot2}
\end{equation}
where $\rho(0)$ is the central ({\bf r} = 0) value of the 
density distribution $\rho({\bf r})$. As a typical
example (for $\rho \simeq \rho(0) \simeq \rho_{sat}$) 
one finds a sequence of coefficients $U_n$ as shown in Fig.~\ref{figU}.
Note once again that the leading attractive piece $U_3$ 
(proportional to the density $\rho$) does not receive 
contributions from the individually much stronger background 
self-energies $\Sigma_V^{(0)} \simeq  -\Sigma_S^{(0)} \simeq 0.35$ 
GeV which cancel in the energy per particle but coherently 
build up the large spin-orbit coupling in finite nuclei. 
The terms of order $\rho^{4/3}$ and $\rho^{5/3}$ 
(with coefficients $U_4$ and $U_5$) are repulsive and 
successively smaller, while the "three-body" term $U_6$ 
proportional to $\rho^2$ is attractive and even 
smaller in magnitude.

It might appear that the attractive $\rho^2$ term causes 
nuclear matter to collapse at very high density. We have 
checked that stability persists up to at least three times 
the density of normal nuclear matter. At even higher densities, 
expansions such as~(\ref{sppot}) must be carried to higher orders
in $k_f$. One should recall, however, that the limit of 
applicability for the present chiral approach  is imposed 
by the condition that $k_f$ must remain sufficiently 
small compared to the reference scale $\Lambda$.

\begin{figure}
\includegraphics[scale=0.55,angle=0]{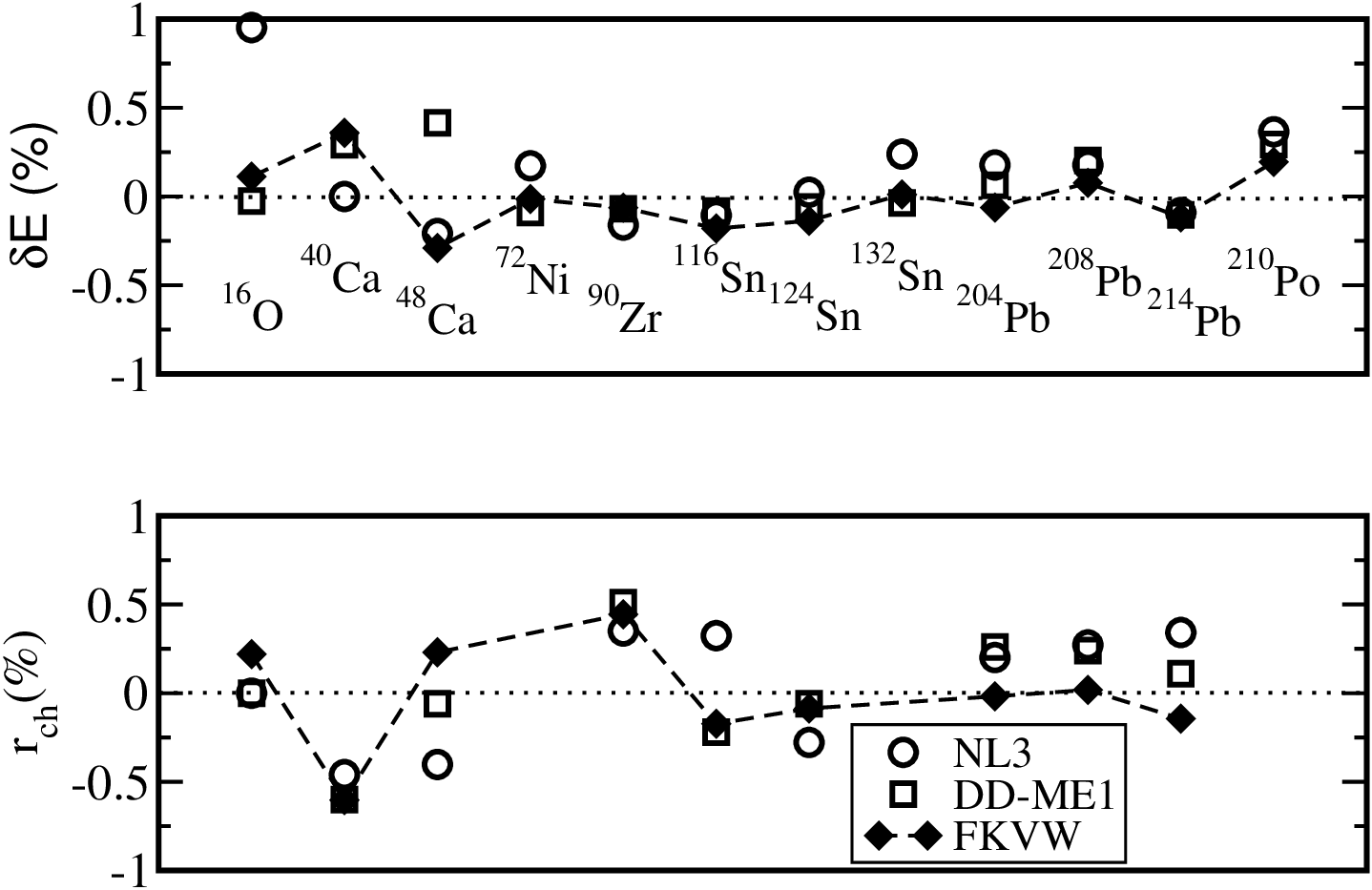}
\vspace{0.5 cm}
\caption{\label{figE}
Deviations (in percent) of the calculated binding energies (upper panel)
and charge radii (lower panel) from the experimental 
values~\cite{Audi2003,NMG.94}, 
for three relativistic effective interactions: the
non-linear meson-exchange NL3~\cite{NL3} (circles), the density-dependent 
meson-exchange DD-ME1~\cite{Nik.02} (squares), and the microscopic 
density-dependent point-coupling 
interaction FKVW (diamonds). The calculations are performed in the 
RMF+BCS model with empirical gaps \cite{Mo.Ni}.
}
\end{figure}

Using the input parameters specified earlier in this section, 
let us have a look at a first series of results achieved 
in comparison with data. In Fig.~\ref{figE} we plot the 
relative deviations between  calculated and experimental values, 
$((\mathcal{O}^i_{th} - \mathcal{O}^i_{exp})/\mathcal{O}^i_{exp})$ in percent, 
of binding energies and charge radii for all nuclei that 
have been included in the fit (see Tab.~\ref{tab2}). The 
RMF+BCS results obtained with
the FKVW parameter set are compared to those of two standard 
phenomenological relativistic mean-field models using a) the  
non-linear meson exchange interaction NL3~\cite{NL3}, and b)
the density-dependent meson-exchange interaction DD-ME1~\cite{Nik.02}. 
The binding energies and charge radii calculated in the present work 
are comparable or superior to those produced with two of the 
most accurate phenomenological relativistic mean-field interactions
(the number of free parameters is seven for NL3, and eight for DD-ME1). 

\section{\label{secIV}Ground-state properties of spherical and
deformed nuclei}

\begin{figure}
\includegraphics[scale=0.55,angle=0]{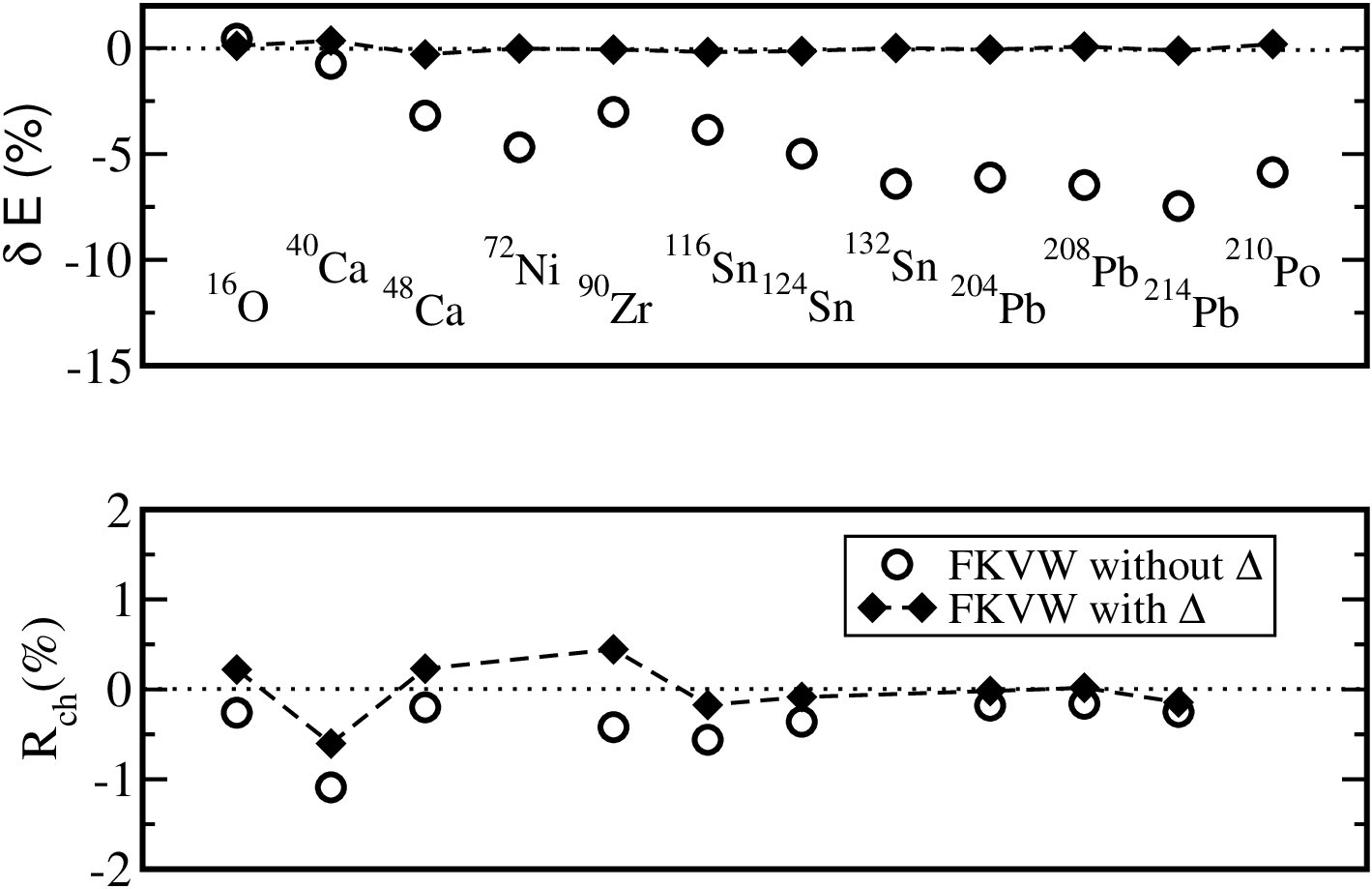}
\vspace{0.5 cm}
\caption{\label{noDelta}
Deviations (in percent) of the calculated binding energies (upper panel)
and charge radii (lower panel) from the experimental 
values~\cite{Audi2003,NMG.94} for the density-dependent 
point-coupling interaction FKVW with explicit inclusion of $\Delta (1232)$
excitations (diamonds) used in the present work, and for the previous 
version~\cite{Fi.03} which did not include the $\Delta (1232)$ degree 
of freedom (cirles).
}
\end{figure}

In Ref.~\cite{Fi.03} a first version of our relativistic 
point-coupling model (at that stage without explicit 
inclusion of $\Delta(1232)$ degrees of freedom)
has been tested in an analysis of
the equations of state for symmetric and asymmetric nuclear matter,
and of bulk and single-nucleon properties of light and
medium heavy $N \approx Z$ nuclei. The detailed isospin 
dependence of the effective interaction was, however, 
less than perfect. For heavier $N > Z$ nuclei such as 
$^{208}Pb$ the calculated binding energies were deviating 
considerably (by more than 5 \%) from the experimental 
values. In the present version of the model, 
both the isoscalar and the isovector channels of the 
exchange-correlation 
functional incorporate the additional contribution of two-pion exchange 
with single and double virtual $\Delta$(1232)-isobar excitations. 
Much better isospin properties of nuclear matter have thus been obtained. 
One can therefore expect improved results also for ground-state properties 
of $N\neq Z$ nuclei calculated with the advanced nuclear energy density 
functional derived the present work. This is illustrated in 
Fig.~\ref{noDelta} where for the set of spherical nuclei that were used in 
the fit (see Tab.~\ref{tab2}), we compare the deviations between calculated
and experimental binding energies and charge radii, respectively, for the
FKVW interaction with explicit inclusion of $\Delta$(1232) excitations, and 
for the previous version of the interaction \cite{Fi.03} which did not 
include the $\Delta$(1232) degree of freedom. The latter displays a 
systematic underbinding with the increase of the neutron -- proton asymmetry
in heavier nuclei, and the deviations exceed 5\% already for Sn isotopes.
This is corrected with the additional contribution of the 
$\Delta$(1232)-isobar, and the new effective interaction reproduces with 
high accuracy the empirical masses. We also notice an improvement for the
calculated charge radii (lower panel of Fig.~\ref{noDelta}), although not
as dramatic as in the case of binding energies.

In this section the improved effective FKVW interaction will be tested in 
self-consistent calculations of ground-state observables for spherical 
and deformed medium-heavy and heavy nuclei. The calculations, including
open-shell nuclei, are performed in the framework of the relativistic 
Hartree-Bogoliubov (RHB) model, a relativistic
extension of the conventional Hartree-Fock-Bogoliubov method,
that provides a basis for a consistent microscopic 
description of ground-state properties of  
medium-heavy and heavy nuclei, low-energy excited states, 
small-amplitude vibrations, and 
reliable extrapolations toward the drip lines \cite{VALR.05}. 
In the particle-hole channel we employ the new microscopic
FKVW interaction. In comparison we also 
present calculations, for spherical nuclei, using the DD-ME1 interaction.
The DD-ME1 model Lagrangian is
characterized by a phenomenological density dependence of the $\sigma$,
$\omega$ and $\rho$ boson-nucleon vertex functions, adjusted to properties
of nuclear matter and finite nuclei. The DD-ME1 effective interaction has 
recently been employed in a number of studies of ground-state properties 
of spherical and deformed nuclei \cite{Nik.02,NVLR.04}, and of 
multipole giant resonances in spherical nuclei \cite{NVR.02,VNR.03,PNVR.04}.
When compared to results obtained with 
standard nonlinear relativistic mean-field effective forces, 
the DD-ME1 interaction gives an improved description of
asymmetric nuclear matter and of ground-state properties of $N \neq Z$ nuclei.
In the relativistic random phase approximation (RRPA), the 
DD-ME1 effective interaction reproduces the experimental excitation
energies of multipole giant resonances in spherical nuclei.

Pairing effects in nuclei are restricted to a narrow window of a few MeV 
around the Fermi level. Their scale
is well separated from the scale of binding energies which
are in the range of several hundred to thousand MeV. 
There is no empirical evidence for any
relativistic effect in the nuclear pairing field.
Therefore pairing can be treated as a non-relativistic phenomenon \cite{Mil}.
In most applications of the RHB model the pairing part of the well 
known and successful Gogny force~\cite{BGG.84} has been used in the
particle-particle channel. We will conveniently proceed in the same 
way, noting at the same time that the chiral N-N potential produces 
very similar pairing matrix elements \cite{Nor_gap}.

\subsection{\label{secIVb}Spherical nuclei}

\begin{figure}
\includegraphics[scale=0.55,angle=0]{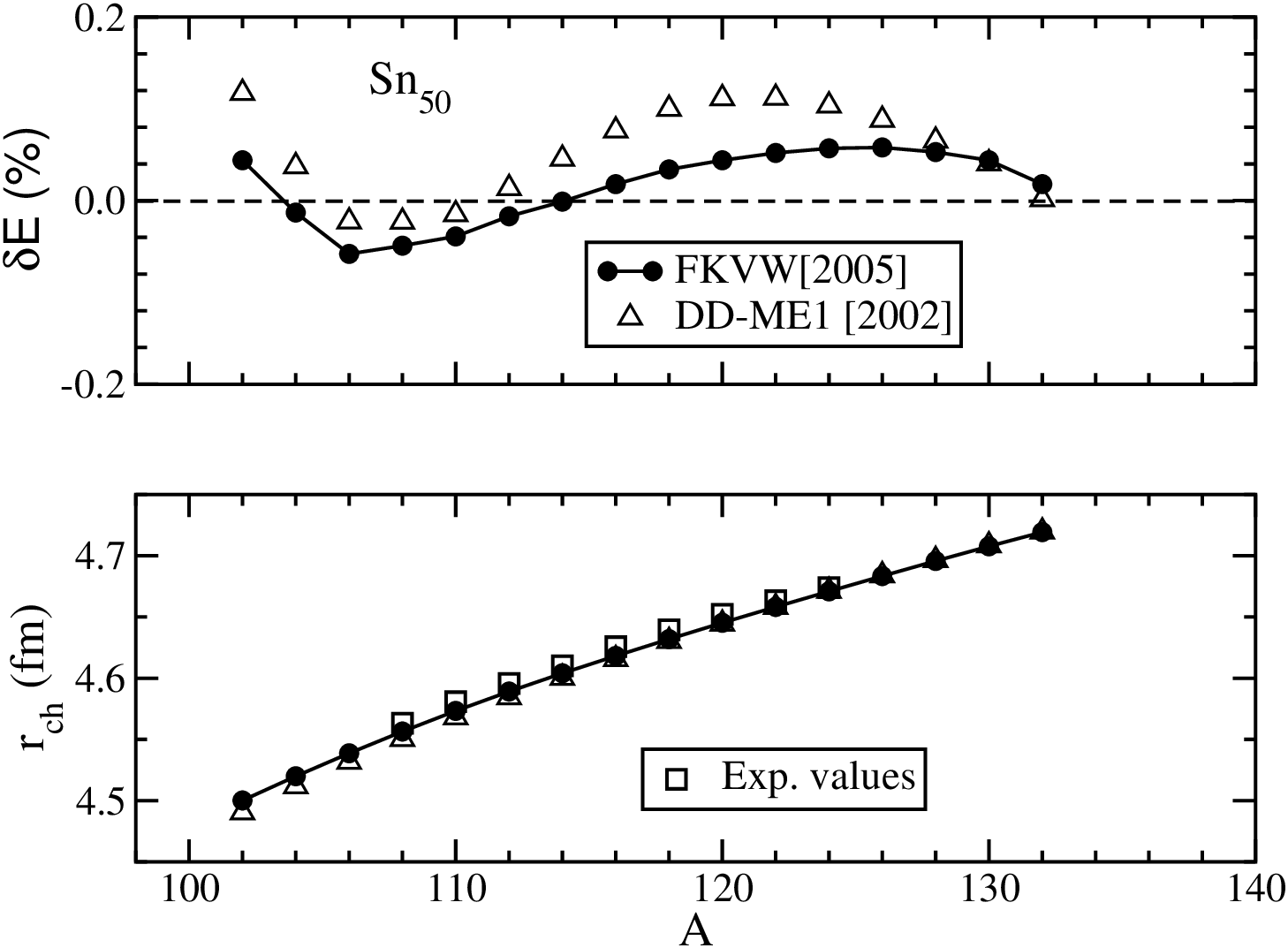}
\vspace{0.5 cm}
\caption{\label{tin} The deviations (in percent) of the calculated binding 
energies from the experimental values (upper panel) ~\cite{Audi2003}, 
and the calculated charge
radii in comparison with data~\cite{NMG.94}, for the chain 
of even-A $Sn$ isotopes. The theoretical 
values are calculated in the RHB model with the DD-ME1~\cite{Nik.02} 
(triangles) and FKVW (dots) density-dependent effective interactions, 
and with the Gogny interaction in the pairing channel.
}
\end{figure}

In order to test the detailed isospin dependence 
we investigate now the systematics of isotopic chains. 
Deviations (in percent) of calculated binding energies 
from their experimental values \cite{Audi2003} are shown 
in Fig.~\ref{tin} for a series of even-A $Sn$ isotopes. 
In addition to the FKVW interaction,
computations are performed with phenomenological density-dependent 
interaction DD-ME1. The Gogny D1S force is used 
for the pairing interaction. Both with FKVW and DD-ME1, 
very good results are found over the entire
major shell $50 \leq N \leq 82$. For the new microscopic FKVW 
interaction in particular, the absolute deviations
of the calculated masses from data do not exceed 0.1 \%. In the lower 
panel of Fig.~\ref{tin} we display the calculated 
charge radii of $Sn$ isotopes in comparison with the experimental 
values~\cite{NMG.94}. For both interactions the theoretical values are in 
excellent agreement with data. 

\begin{figure}
\includegraphics[scale=0.55,angle=0]{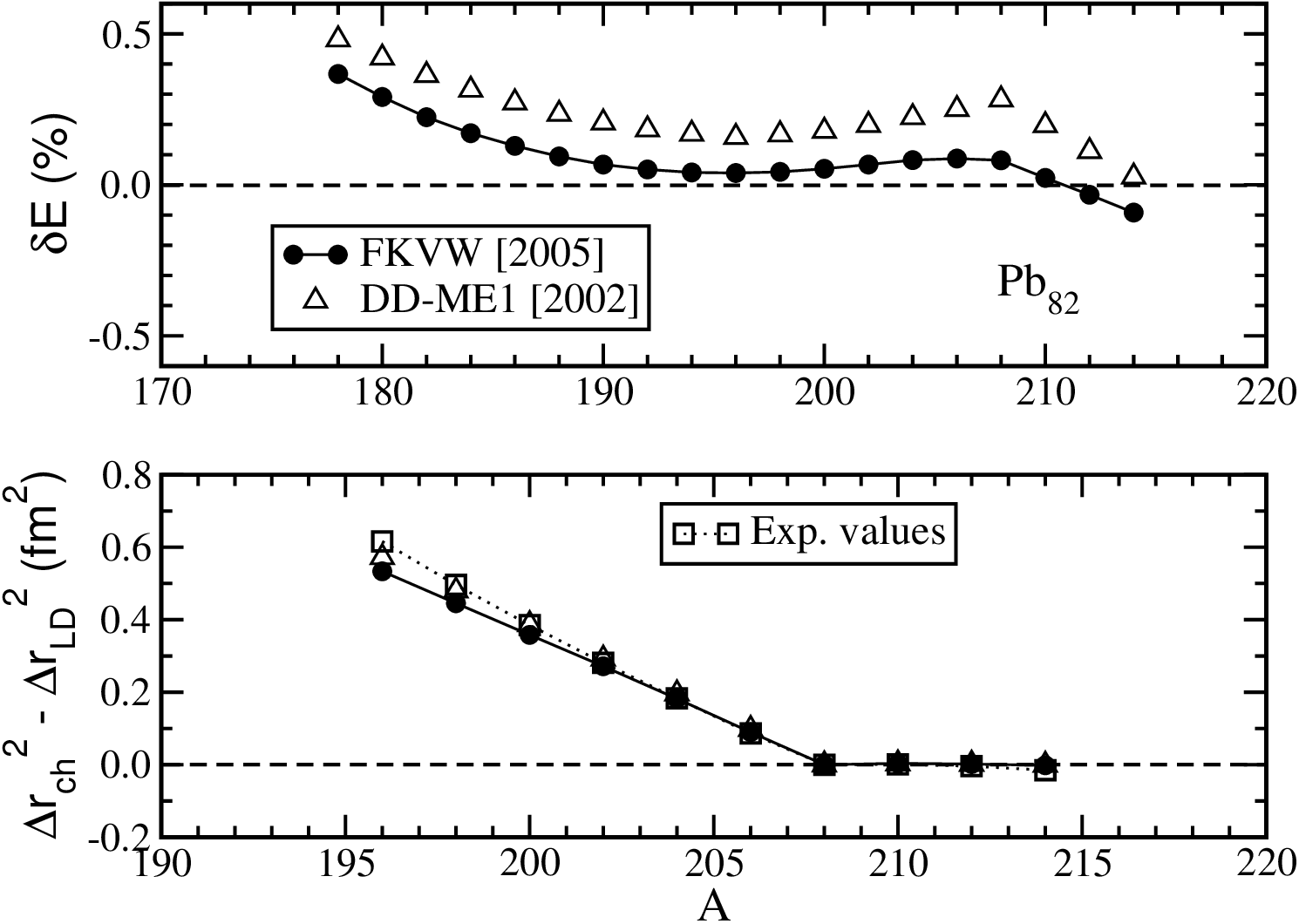}
\vspace{0.5 cm}
\caption{\label{lead}
The deviations (in percent) of the calculated binding 
energies from the experimental values (upper panel)~\cite{Audi2003}, 
and the calculated charge
isotope shifts in comparison with data~\cite{Rad}, 
for the chain of even-A $Pb$ isotopes. 
The charge isotope shifts are defined: 
$\Delta r_{ch}^2 = r_{ch}^2(A) - r_{ch}^2(^{208}Pb)$ and 
$ \Delta r_{LD}^2 = r_{LD}^2(A) - r_{LD}^2(^{208}Pb)$, where
the liquid-drop estimate is $r^2_{LD}(A) = {3\over 5}r_0^2A^{2/3}$.
The theoretical values are calculated in the RHB model with the DD-ME1 
(triangles)~\cite{Nik.02}and 
FKVW (dots) density-dependent effective interactions, 
and with the Gogny interaction in the pairing channel.
}
\end{figure}

The isotopic dependence of the 
deviations (in percent) between the calculated binding 
energies and the experimental values for even-A $Pb$ nuclei~\cite{Audi2003}, 
is plotted in the upper panel of Fig.~\ref{lead}. It is 
interesting to note that, although DD-ME1 and FKVW represent 
different physical models, they display a 
similar mass dependence of the calculated binding energies
for the Pb isotopic chain. On a quantitative level the FKVW 
interaction produces better results, with the absolute deviations
of the calculated masses below 0.1 \% for $A \geq 190$. In lighter
Pb isotopes one expects that the observed shape coexistence phenomena 
will have a pronounced effect on the measured masses. 

Because of the intrinsic isospin independence of the effective
single-nucleon spin-orbit potential, relativistic mean-field
models naturally reproduce the anomalous charge isotope
shifts~\cite{SLR.93}. The well known example
of the anomalous kink in the charge isotope shifts of
$Pb$ isotopes is illustrated in the lower panel of Fig.~\ref{lead}. 
The results of RHB calculations with the DD-ME1 and FKVW effective 
interactions are shown in comparison with experimental values \cite{Rad}.
Both interactions reproduce in detail the A-dependence
of the isotope shifts and the kink at $^{208}Pb$.
 
\begin{figure}
\includegraphics[scale=0.55,angle=0]{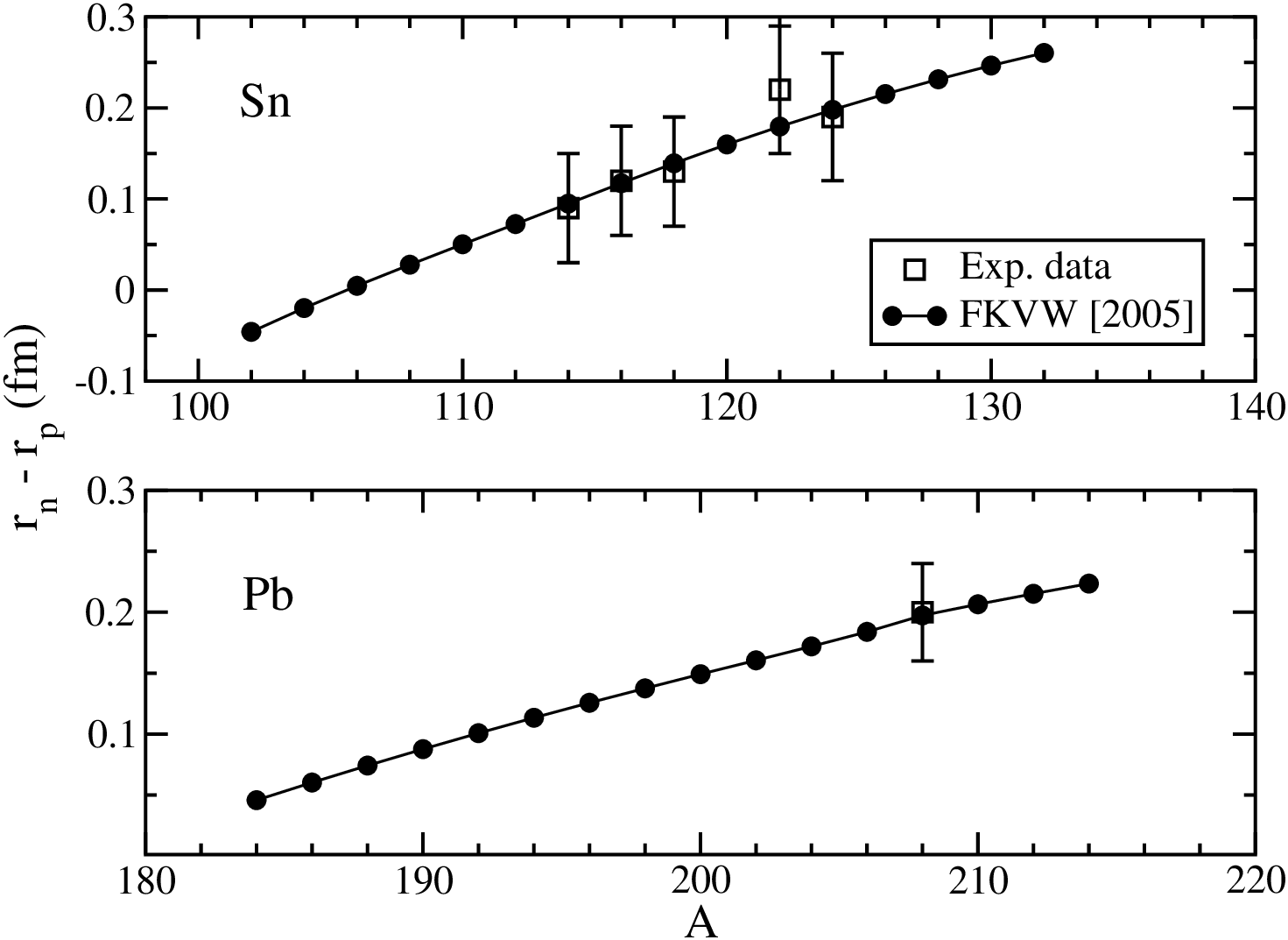}
\vspace{0.5 cm}
\caption{\label{rn_rp}
FKVW plus Gogny RHB-model predictions for the differences between the neutron
and proton rms radii of $Sn$ (upper panel) and $Pb$ (lower panel) isotopes, 
in comparison with available data~\cite{Kra.99,SH.94,Kra.94}.}
\end{figure}

The differences between the r.m.s. radii of neutron and proton
ground-state distributions of $Sn$ and $Pb$ nuclei are shown in
Fig.~\ref{rn_rp}. Results obtained with 
the FKVW interaction are compared with 
data~\cite{Kra.99} for the $Sn$ isotopes, and with the empirical
value of $r_n - r_p$ in $^{208}Pb$ ($0.20 \pm 0.04$ fm from proton
scattering data~\cite{SH.94}, and $0.19 \pm 0.09$ fm from 
the excitation of the isovector 
giant dipole resonance by $\alpha$-scattering~\cite{Kra.94}).

The determination of neutron density distributions provides not only basic
nuclear structure information, but it also places important
additional constraints on the isovector channel of the 
effective interactions used in nuclear models.
In a recent analysis of neutron
radii in non-relativistic and covariant mean-field models~\cite{Fur.01},
the linear correlation between the neutron skin and the asymmetry energy
was studied. In particular, it has been  shown that there is a
very strong linear correlation between the neutron skin thickness in
$^{208}Pb$ and the parameters that determine the
asymmetry energy. The excellent agreement between the
calculated $r_n - r_p$ and the available data confirms 
that the isovector channel of the microscopic effective interaction 
FKVW is correctly represented. Note that this is 
achieved without ad-hoc introduction of
a phenomenological $\rho$-meson exchange interaction. 

\begin{figure}
\includegraphics[scale=0.55,angle=0]{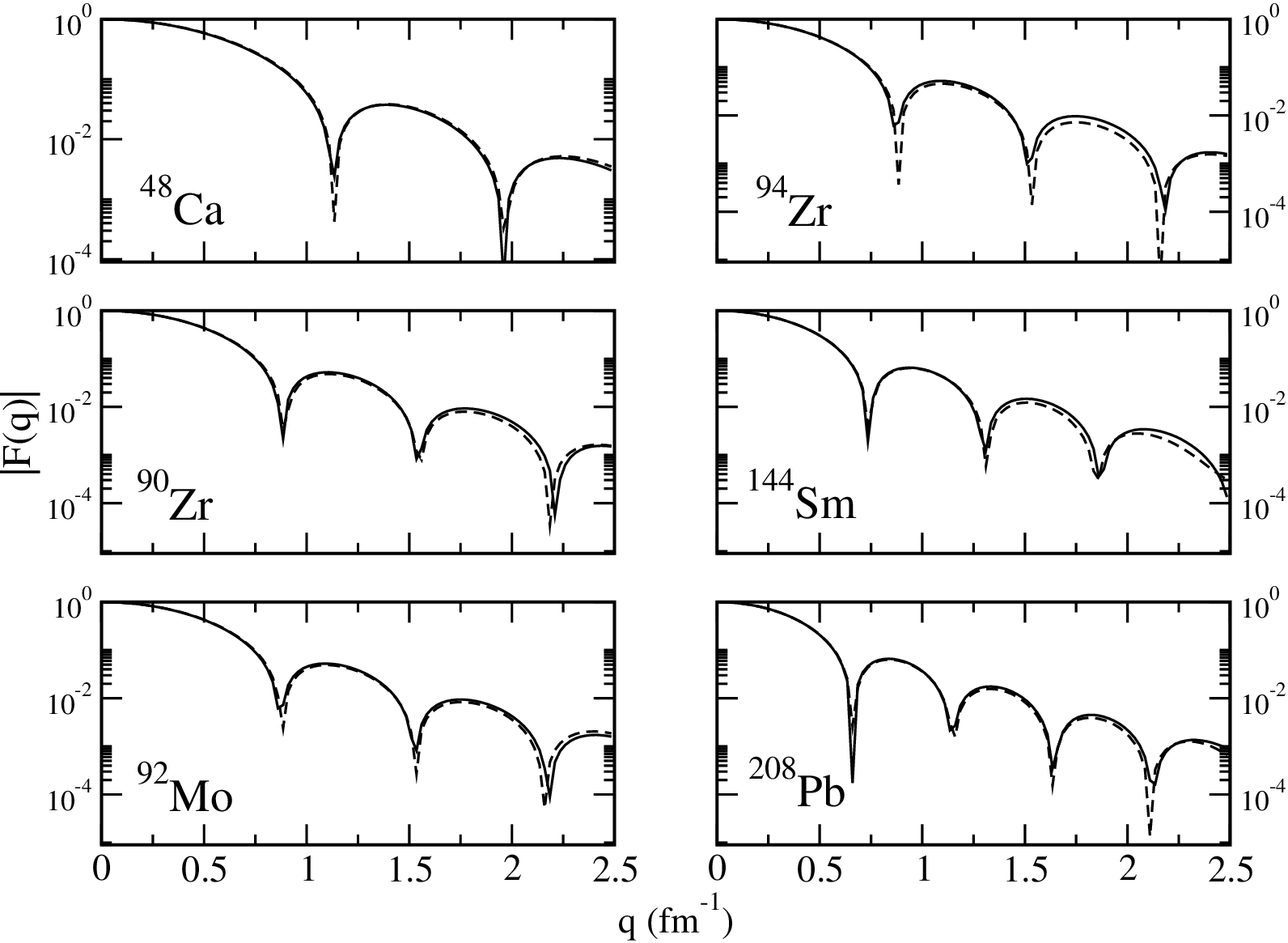}
\vspace{0.5 cm}
\caption{\label{ff}
Charge form factors of $^{48}Ca$, $^{90}Zr$, $^{92}Mo$, $^{94}Zr$,
$^{144}Sm$ and $^{208}Pb$ calculated 
in the relativistic point-coupling model with the FKVW density-dependent 
effective interaction (full lines), in comparison with the experimental 
form factors (dashed lines) \cite{Vri.87}.
}
\end{figure}

The FKVW model is based on the framework of density functional theory 
and, therefore, a natural test is the comparison of the calculated 
density distributions with available data. In Fig. \ref{ff}
we display several calculated charge form factors
in comparison with data \cite{Vri.87}. In addition to 
$^{48}Ca$, $^{90}Zr$, and $^{208}Pb$, for which the charge radii i.e. 
the first minima of the charge form factors have been used in the 
fit of the parameter set FKVW (see Table \ref{tab2}), we also 
plot the charge form factors of $^{92}Mo$, 
$^{94}Zr$, and $^{144}Sm$.
The theoretical charge density distributions are obtained by folding
the point proton density distribution with 
the proton charge distribution of exponential form $\sim e^{-\Lambda r}$
(reflecting a dipole form factor), with the empirical proton charge radius. 
The corresponding form factors
are plotted as functions of the momentum transfer $\bm{q}$:
\begin{equation}
F_{ch}(\bm{q}) = \frac{1}{Z} \rho_{ch}(\bm{q}) \left[ 1 + 
\frac{\bm{q}^2}{8
\langle P^2_{cm}\rangle} + \ldots \right] \; ,
\end{equation}
where $\rho_{ch}(\bm{q})$ is the Fourier transform of the 
spherical charge density and the second term is a
center-of-mass correction. 
Higher-order effects in $\bm{q}^2/P^2_{cm}$ are negligible.
The excellent agreement 
between the calculated and experimental charge form factors 
for momenta $|\bm{q}| \leq 2.5$ fm$^{-1}$ demonstrates that the 
FKVW interaction reproduces not only the moments of the distributions, 
but also the detailed charge density profiles.

\begin{figure}
\includegraphics[scale=0.55,angle=0]{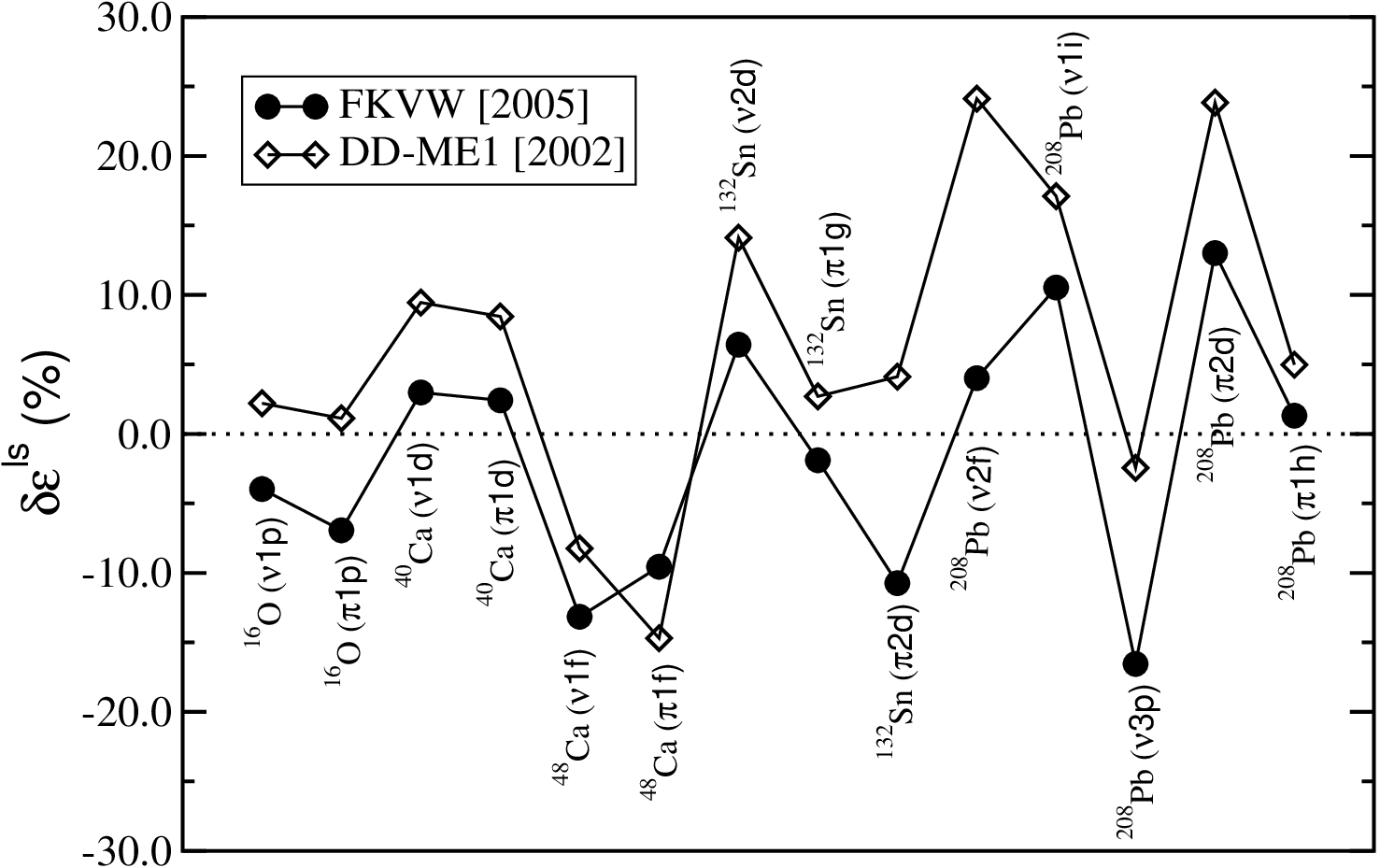}
\vspace{0.5 cm}
\caption{\label{spin_orbit}
The deviations (in percent) between the theoretical and experimental 
values~\cite{NUDAT} 
of the energy spacings between spin-orbit partner-states in doubly 
closed-shell nuclei. The calculated spin-orbit splittings correspond to the 
relativistic density-dependent point-coupling interaction FKVW (dots), 
and the relativistic density-dependent meson-exchange interaction DD-ME1 
(diamonds)~\cite{Nik.02}.}
\end{figure}

One of the principal advantages of using the 
relativistic framework lies in the fact
that the effective single-nucleon spin-orbit potential 
arises naturally from the Dirac equation. 
The single-nucleon potential does not introduce any adjustable parameter for 
the spin-orbit interaction. In the FKVW model, in particular, 
the large effective spin-orbit potential in finite nuclei is generated
by the strong scalar and vector condensate background 
fields of about equal magnitude and
opposite sign, induced by changes of the QCD vacuum in the presence
of baryonic matter \cite{Fi.03}. In Fig. \ref{spin_orbit} we plot
the deviations (in percent) between the calculated and experimental 
values of the energy spacings between spin-orbit partner-states in
a series of doubly closed-shell nuclei. Even though the Kohn-Sham 
orbitals cannot be directly identified with the single-nucleon states
in the interacting system, the calculated energy spacings between
spin-orbit partners, especially the ones close to 
the Fermi surface, compare well with the experimental values. 
The experimental data are 
from Ref.~\cite{NUDAT}, and the theoretical spin-orbit splittings 
have been calculated with the FKVW and DD-ME1 interactions. 
For the phenomenological DD-ME1 interaction 
the large scalar and vector nucleon self-energies which generate 
the spin-orbit potential, arise from the exchange of ``sigma'' and
``omega'' bosons with adjustable strength parameters.
We notice that, even though the values calculated with DD-ME1 are already
in very good agreement with experimental data, a further improvement
is obtained with the FKVW interaction. This remarkable agreement  
indicates that the initial estimates 
Eqs. (\ref{estim1}) and (\ref{estim2}) for the condensate background 
couplings have perhaps been more realistic than anticipated,  
considering the uncertainties of lowest-order in-medium QCD sum rules. 

\subsection{\label{secIVc}Deformed nuclei}

Deformed nuclei with $N > Z$ present
further important tests for nuclear structure models. 
Ground-state properties, in particular, are sensitive 
to the isovector channel of the effective interaction, 
to the spin-orbit term of the effective 
single-nucleon potentials and to the effective mass.

In this section we test our advanced model 
in the region $60\le Z \le80$. We compare 
predictions of the RHB calculations for the total binding energies, 
charge radii and ground-states quadrupole deformations of nine 
even-Z isotopic chains with available data. 
The FKVW effective interaction is used in the particle-hole
channel, and pairing correlations are described by the pairing part of the
finite range Gogny D1S interaction~\cite{BGG.84}.

\begin{figure}
\includegraphics[scale=0.55,angle=0]{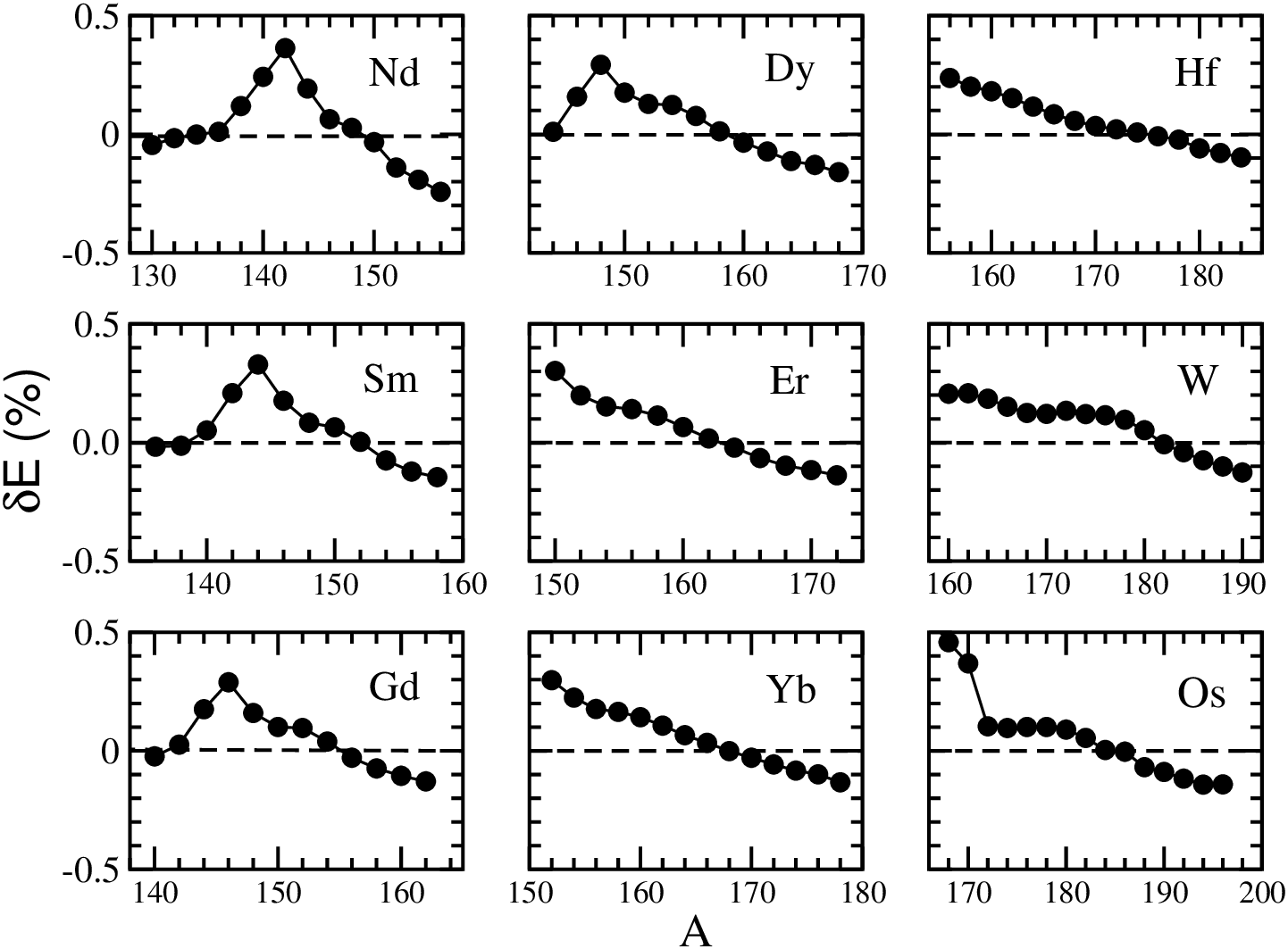}
\vspace{0.5 cm}
\caption{\label{E_def}
The deviations (in percent) of the calculated binding 
energies from the experimental values~\cite{Audi2003} of 
$Nd$, $Sm$, $Gd$, $Dy$, $Er$, $Yb$, $Hf$, $W$, and $Os$ isotopes. The 
ground-state binding energies have been 
calculated in the RHB model with the FKVW parameterization, 
and with the Gogny interaction in the pairing channel.
}
\end{figure}

The deviations (in percent) of the calculated binding 
energies from the experimental data~\cite{Audi2003} for
$Nd$, $Sm$, $Gd$, $Dy$, $Er$, $Yb$, $Hf$, $W$, and $Os$ isotopes
are plotted in Fig.~\ref{E_def}. Good agreement is found over
the entire region of deformed nuclei. The maximum 
deviation of the calculated binding
energies from data is below $0.5$\% for all isotopes.

\begin{figure}
\includegraphics[scale=0.55,angle=0]{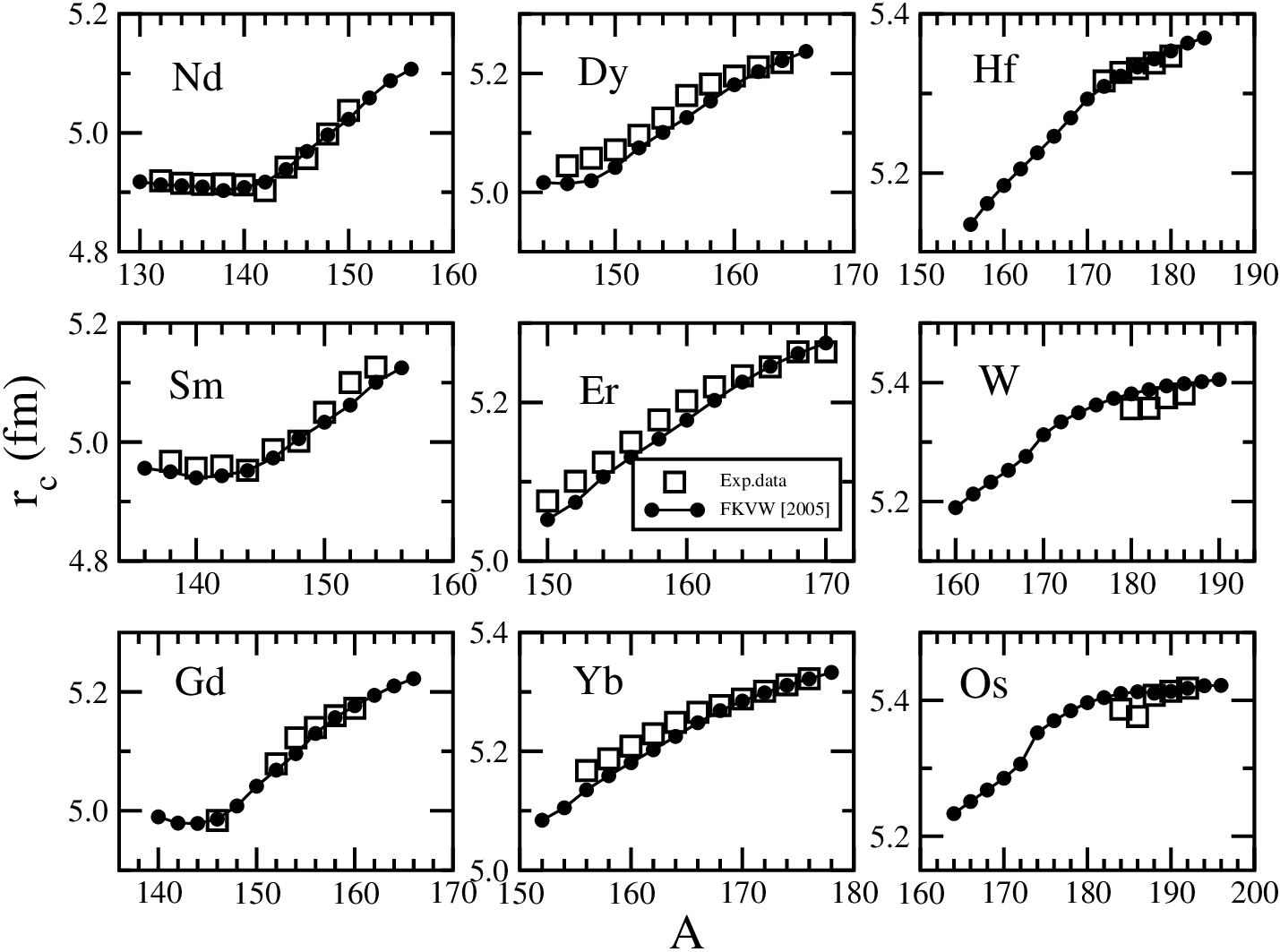}
\vspace{0.5 cm}
\caption{\label{RC_def}
RHB model (FKVW interaction plus Gogny pairing) predictions for the 
ground-state charge radii of  
$Nd$, $Sm$, $Gd$, $Dy$, $Er$, $Yb$, $Hf$, $W$, and $Os$ isotopes, 
in comparison with available data \cite{NMG.94}.}
\end{figure}

In Fig.~\ref{RC_def} we compare the calculated charge radii
with data from Ref.~\cite{NMG.94}. The charge density is
constructed by folding the calculated (point) 
proton density distribution with the empirical proton-charge 
distribution. The calculation of the charge $rms$ radius from the 
full charge form factor is quite involved for deformed 
nuclei \cite{BHR.03,BFHN}. We employ the frequently used simplified
expression
\begin{equation}
r_c = \sqrt{r_p^2+0.64~{\rm fm}^2} \;,
\end{equation}
where $r_p^2$ is the nuclear mean-square radius of the 
(point) proton density distribution.
The calculated charge radii reproduce in detail 
the experimental isotopic trends.

\begin{figure}
\includegraphics[scale=0.55,angle=0]{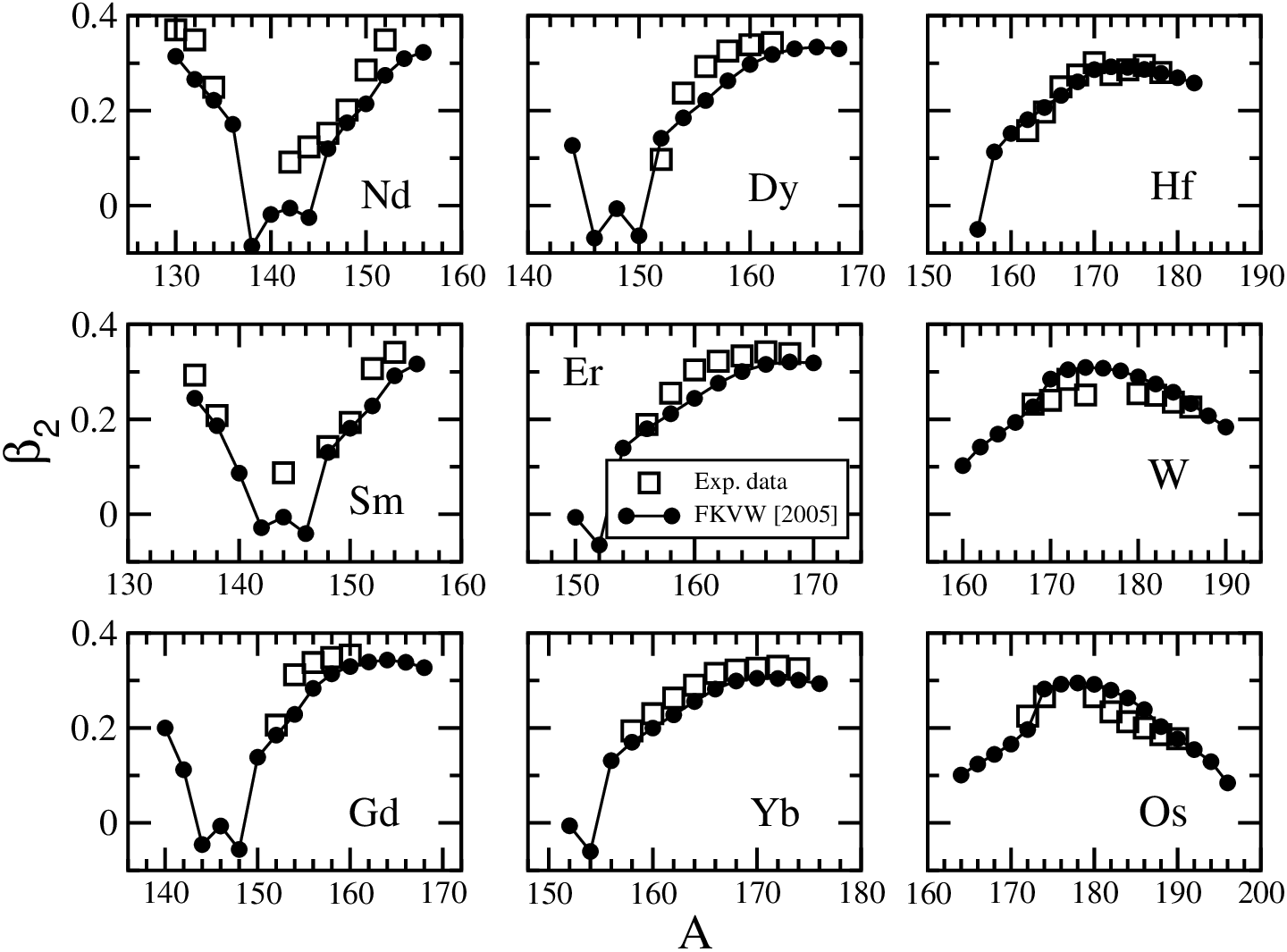}
\vspace{0.5 cm}
\caption{\label{Beta_def}
Comparison between the RHB model (FKVW interaction plus Gogny pairing) 
predictions for the ground-state
quadrupole deformation parameters of the $Nd$, $Sm$, $Gd$, 
$Dy$, $Er$, $Yb$, $Hf$, $W$, and $Os$
isotopes, and experimental
values \cite{RNT.01}.
}
\end{figure}

The ground-state quadrupole deformation parameter $\beta_2$, 
proportional to the expectation value of the quadrupole 
operator $\langle \phi_0| 3 z^2 - r^2 |\phi_0 \rangle$,
is calculated according to the prescription of Ref.~\cite{LQ.82}.
The theoretical values of the quadrupole deformation parameter are 
displayed in Fig.~\ref{Beta_def}, 
in comparison with the empirical data extracted 
from $B(E2)$ transitions ~\cite{RNT.01}.
We notice that the RHB results reproduce not only the 
global trend of the data but also the saturation of quadrupole deformations
for heavier isotopes.

The structure of nuclei far from stability provides a particularly 
important testing ground for global effective interactions. 
An interesting question, therefore, is how far from stability 
can we extrapolate the predictions of the FKVW interaction. 
While an extensive 
analysis of this problem is beyond the scope of the present 
work, we consider in detail the example of neutron-deficient $Pb$ 
isotopes which exhibit an interesting variety of coexisting shapes. They are
characterized by the competition of low-lying prolate and oblate minima.
Recent data~\cite{Shape_exp} on energy spectra and charge radii 
indicate that, although the ground states of neutron-deficient $Pb$ nuclei 
are spherical, both oblate 
(two-particle -- two-hole proton excitations
across the $Z=82$ major shell) and prolate (four-particle -- four-hole proton 
excitations) low-lying minima can be observed for $N < 110$ 
at almost identical excitation energies. 

A number of theoretical analyses of shell 
quenching and shape coexistence phenomena in neutron-deficient $Hg$, $Pb$ 
and $Po$ nuclei has been reported in recent years. The quantitative
description of coexisting spherical, oblate and prolate
minima necessitates restoration of broken symmetries and the 
explicit treatment of quadrupole fluctuations. In particular, 
excellent results for shape coexistence
in neutron-deficient $Pb$ isotopes have recently been obtained by 
performing configuration mixing of angular momentum projected self-consistent
mean-field states, calculated with the finite range and density dependent
Gogny interaction \cite{RER.04}, and with the Skyrme interaction SLy6 
supplemented by a density-dependent zero-range pairing force~\cite{GCM}.

The model employed in the present work does not include angular 
momentum projection, nor can it account for configuration mixing effects. 
Nevertheless, we can compare the potential 
energy curves, calculated as functions of the axial quadrupole
deformation, with the corresponding mean-field energy curves calculated
with the Gogny \cite{RER.04} and Skyrme \cite{GCM} interactions.

\begin{figure}
\includegraphics[scale=0.55,angle=0]{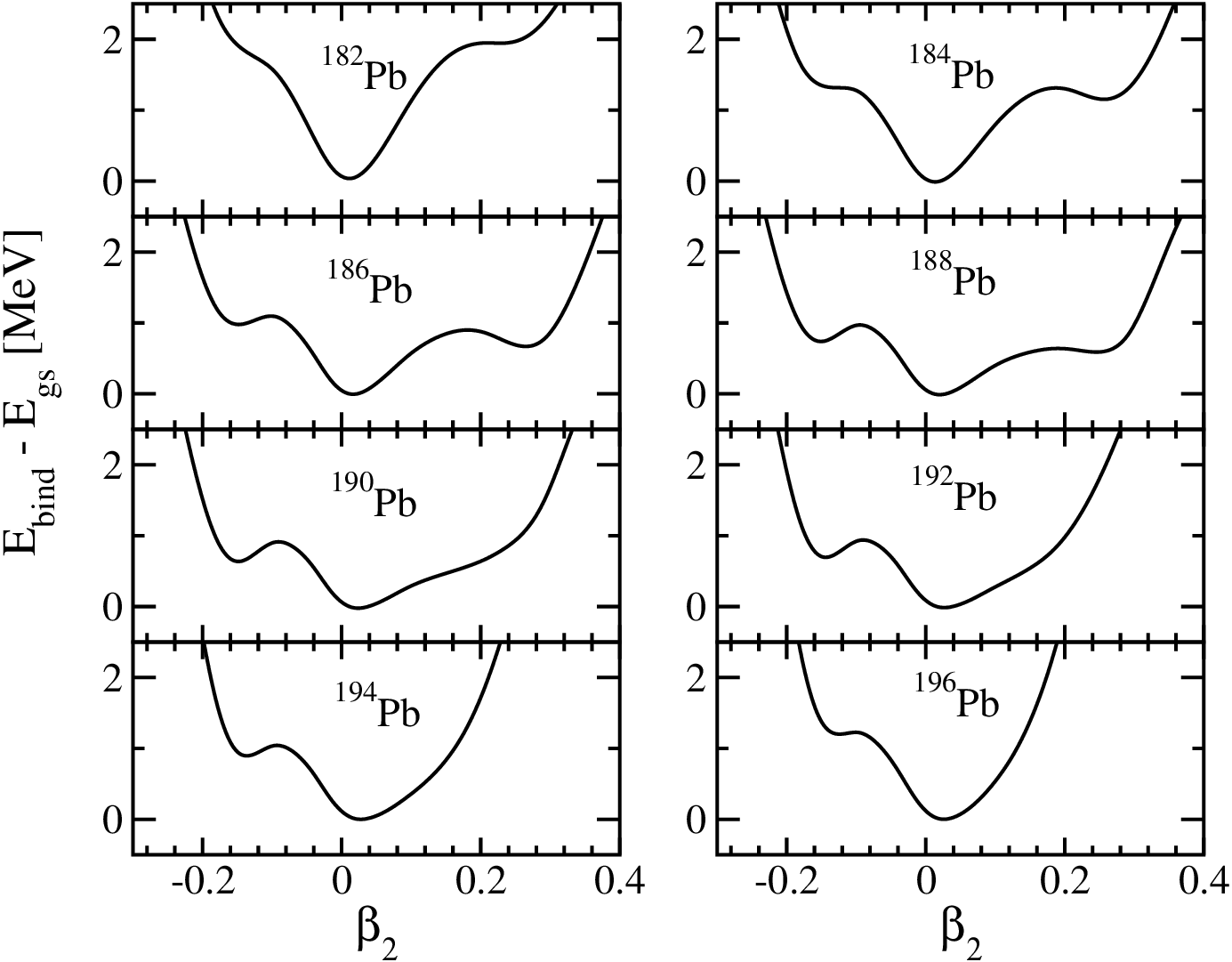}
\vspace{0.5 cm}
\caption{\label{Shape}
Binding energy curves of even-A $Pb$ isotopes as functions of the quadrupole 
deformation $\beta_2$. The curves correspond to RHB 
calculations with constrained 
quadrupole deformation.}
\end{figure}

In Fig.~\ref{Shape} we display the calculated binding energy curves of even-A 
$^{182-196}Pb$ isotopes as functions of the quadrupole deformation. The
curves correspond to axially deformed RHB model solutions with constrained
quadrupole deformation. 
The effective interaction is FKVW + Gogny D1S (pairing).
The general agreement with the mean-field potential energy 
curves of Refs.\cite{RER.04} and \cite{GCM} is satisfactory and, in 
particular, the present calculation predicts 
the coexistence of oblate and prolate minima at almost identical
excitation energies in $^{186,188}Pb$, in agreement with data.
The relative excitation energies of coexisting minima based on 
different intruder configurations are determined by the 
spherical magic and semi-magic energy gaps. For $A \geq 190$ the 
minimum on the prolate side vanishes, in accordance with the 
nonrelativistic calculations performed with the Gogny and 
Skyrme-SLy6 interactions. In these nuclei, however, we also 
calculate a tiny shift of the
ground-state minimum toward prolate deformation, a result not 
predicted by the two non-relativistic interactions, and not 
corroborated by available data. Although the quantitative analysis
of the structure of neutron-deficient $Pb$ nuclei must include 
additional correlations related to the restoration of broken 
symmetries and quadrupole fluctuations, it is indeed an encouraging 
result that the FKVW interaction 
reproduces the isotopic trend of the coexistence of shapes 
in heavy nuclei far from stability. This reinforces the 
confidence in the detailed
behavior of the isospin-dependent interaction derived from 
chiral two-pion exchange
dynamics, improved by explicit $\Delta(1232)$ components.

\section{\label{secV}Summary, conclusions and further comments}

We have derived a relativistic nuclear energy density functional
with connections to two closely linked features of QCD in the 
low-energy limit: \\a) in-medium changes of vacuum condensates;
\\ b) spontaneous chiral symmetry breaking. 

The leading changes of the chiral (quark) condensate and quark density 
in the presence of baryonic matter are sources of strong (attractive) 
scalar and (repulsive) vector fields experienced by nucleons in the 
nucleus. These fields produce Hartree potentials of about 0.35 GeV 
in magnitude at nuclear matter density, in accordance with QCD sum 
rules. While these scalar and vector potentials cancel approximately 
in their contribution to the energy, they are at the origin of the 
large spin-orbit splitting in nuclei. 

The spontaneously broken chiral symmetry in QCD introduces pions as 
Goldstone bosons with well-defined (derivative) couplings to baryons 
plus symmetry breaking corrections. In the present approach the 
exchange-correlation part of the energy density functional is deduced 
from the long- and intermediate-range interactions generated by one- 
and two-pion exchange processes. They have been computed using in-medium 
chiral perturbation theory with explicit inclusion
of $\Delta(1232)$ degrees of freedom which turn out to be important. 
Regularization dependent 
contributions to the energy density, calculated at three-loop level, 
are absorbed in contact interactions
with constants representing unresolved short-distance dynamics. 

This framework is translated into a point-coupling model with 
density-dependent interaction vertices. This is done for the practical 
purpose of deriving and solving self-consistent Dirac equations 
(relativistic analogues of  Kohn-Sham equations) in order to 
determine the nucleon densities which enter the energy functional. 
We demonstrate that this scheme works extremely well when confronted 
with a large number of high-precision nuclear
data over a broad range of spherical and deformed nuclei. 
In fact the quantitative accuracy of the calculated binding 
energies and radii is such that deviations from empirical data are 
generally less than 0.5\% throughout the nuclear chart
and usually much less (0.1\%) for medium-heavy nuclei. It has to be 
noted, however, that these results still do not reach the accuracy of the best 
phenomenological mass tables (recent trends in the determination
of nuclear masses have been reviewed in Ref.~\cite{LPT.03}). Modern 
Skyrme-based microscopic mass formulas, with a total of $\approx 20$ 
empirical parameters, fit the measured masses of more than two 
thousand nuclei with an rms error of less than 700 keV. In addition, 
they also produce excellent results for charge radii, with an rms 
deviation of $\approx 0.025$ fm for the absolute charge radii and charge 
isotope shifts of more than 500 nuclei. The nuclear energy density functional 
developed in this work is not, of course, a mass formula and, with only a 
small set of adjustable parameters, is not designed to compete with such a level of 
accuracy. Nevertheless, it provides an excellent microscopic framework on which
a more fundamental approach to the calculation of nuclear masses can be based.

A special focus in this work is on systematic studies along isotopic chains. 
Binding energies, as well as proton and neutron radii of such systems, 
with fixed number of protons but varying number of neutrons, are an 
excellent testing ground for the detailed isospin properties of the 
effective interaction. The quality of the results points to the fact 
that chiral pion dynamics, and especially the Van der Waals-like two-pion exchange forces 
with inclusion of the $\Delta(1232)$, is capable of generating the proper
isospin dependence required by observables in isotopic sequences.

The construction of the density functional involves an expansion of 
nucleon self-energies in powers of the Fermi momentum up to and including 
terms of order $k_f^6$, or equivalently, $O(\rho^2)$ in the proton and 
neutron densities. Up to this order the present model has seven parameters, 
four of which are related to contact (counter) terms that appear in the 
chiral perturbation theory treatment of nuclear matter. One parameter 
fixes a surface (derivative) term and two more represent the strengths of 
scalar and vector Hartree fields. 

In the "best fit" set which reproduces a large amount of data on nuclear 
ground state properties, five of those seven parameters turn out to be 
surprisingly close to estimates and predictions from in-medium QCD sum 
rules and ChPT calculations for nuclear matter. In particular,
the strong scalar and vector (Hartree) fields required to 
reproduce spin-orbit splittings are remarkably
compatible with the leading-order QCD sum rule estimates. 
The fitted coefficient $b_3$ multiplying the
$O(k_f^3)$ term in the chiral expansion of the nucleon self-energy,  
is within less than 5 \% of the
one from the ChPT calculation for nuclear matter. For the corresponding 
coefficient $a_3$ in the expansion of the isospin dependent nucleon 
self-energy, the deviation (between fit to the nuclear data base and ChPT 
evaluation of the symmetry energy) is even less than 2 \%. The surface term 
required to reproduce the detailed systematics of nuclear radii is within 
less than 10 \% of the ChPT prediction for slightly inhomogeneous nuclear 
matter. This prediction is actually parameter-free since the corresponding 
loop integrals involve only finite, regularization independent pieces. The 
exceptions that deviate from this overall pattern are the two parameters 
related to the three-body contact interaction terms (of order $k_f^6$, 
or $\rho^2$, in the self-energies) for which the nuclear data require more 
attraction than anticipated by the in-medium ChPT estimate. On absolute 
scales, however, these three-body terms are relatively small within the 
hierarchy of the $k_f$ expansion: an encouraging result although it cannot 
strictly be taken as proof of convergence.

In this work we have only considered the low-energy QCD constraints on 
the effective interaction in the particle-hole channel, whereas the pairing 
part of the Gogny D1S force has been used for the effective interaction 
in the particle-particle channel. One could, of course, extend the chiral 
EFT approach to the pairing channel, and this would introduce additional
contact terms. The first step in this direction has been taken in 
Ref.~\cite{Nor_gap}, where an improved version of the chiral nucleon-nucleon
potential at next-to-next-to-leading order has been used to calculate 
the $^1\!S_0$ pairing gap in isospin-symmetric nuclear matter. The  
pairing potential consists of long-range one- and two-pion exchange terms, 
and two NN-contact couplings. It has been shown that the inclusion of the 
two-pion exchange at the next-to-next-to-leading order reduces substantially 
the cut-off dependence of the $^1\!S_0$ pairing gap, indicating reasonable 
convergence of the small momentum expansion. The results are very 
close to those obtained with the phenomenological Gogny D1S force.

The exchange-correlation energy functional in nuclear matter, calculated 
by in-medium chiral perturbation theory, has been used in Kohn-Sham 
calculations of finite nuclei by employing a second-order gradient
correction to the local density approximation. Even though the quality
of the results is on the level of the best phenomenological (non-relativistic 
and relativistic) self-consistent mean-field models, 
obviously the goal is to further improve the accuracy of the 
calculated nuclear ground-state energies and 
density distributions all over the periodic table. In the next step 
this will require the development of an accurate generalized gradient 
approximation for the nuclear exchange-correlation energy.   

An important issue will also be the relation with the universal low-momentum 
interaction $V_{low-k}$ \cite{BKS}, deduced from phase-shift equivalent 
nucleon-nucleon interactions. An apparent close correspondence between 
perturbative in-medium chiral dynamics and $V_{low-k}$ in the low-density 
limit has already been pointed out recently in Ref.\,\cite{Fri.04}. 
This is a non-trivial observation, given that the
input to the chiral approach \cite{Fri.04} has been fixed to nuclear 
matter properties and not to NN phase shifts. The applicability of a 
perturbative expansion scheme for nuclear matter with low-momentum 
interactions, which is the basic hypothesis behind our approach, is 
discussed in Ref.\,\cite{BSFN}. Further studies towards foundations 
along these lines are presently being pursued.

This work has demonstrated that chiral effective field theory provides a
consistent microscopic framework in which both the isoscalar and isovector 
channels of a universal nuclear energy density functional can be formulated. 
The present approach to nuclear DFT establishes a fundamental link between 
low-energy QCD and ground-state properties of finite nuclei. From a practical
point of view, a fully microscopic basis of effective nuclear interactions
is especially important for studies of nuclear structure in regions 
far from the valley of $\beta$-stability, where extrapolations of  
phenomenological (non-relativistic and relativistic) models lack  
predictive power.

\section{Acknowledgments}
Helpful discussions with Stefan Fritsch, Tamara Nik{\v{s}}i{\'{c}} and 
Peter Ring are gratefully 
acknowledged. One of the authors (W.W.) thanks Tony Thomas, the Theory Group
at Jefferson Lab, and ECT* in Trento for their hospitality during the 
preparation of this paper. 

\section{\label{App1}Appendix}

Here we present details of the nucleon self-energies from 
in-medium chiral perturbation theory
and their translation into a point-coupling model with 
density-dependent vertices.
In~\cite{Kai.01jx,Kai.01ra,Fri.04} the momentum and 
density-dependent single particle potential of
nucleons in asymmetric nuclear matter:
\begin{equation}
U(p,k_f) - U_I(p,k_f)\tau_3 \delta + \mathcal{O}(\delta^2) 
\quad {\rm with} \quad \delta =
\frac{\rho^n - \rho^p}{\rho^n + \rho^p} \; ,
\end{equation}
has been calculated within two-loop in-medium 
chiral perturbation theory (ChPT).

From the sums
\begin{eqnarray}
\label{sum}
U(p = k_f,k_f) & = & \Sigma_S^{\rm ChPT} (k_f,\rho) + 
\Sigma_V^{\rm ChPT} (k_f,\rho) \\
- U_I(p = k_f,k_f) \delta & = & \Sigma_{TS}^{\rm ChPT} (k_f,\rho) +
\Sigma_{TV}^{\rm ChPT} (k_f,\rho)\nonumber
\end{eqnarray}
of the scalar and vector nucleon self-energies in the isoscalar 
and isovector channels can be extracted at the Fermi surface, 
$p=k_f$. It has been shown that the differences between scalar 
and vector parts of the ChPT self-energies are very small 
(suppressed by factors $M_N^{-2}$) and linearly proportional to $\rho$. 
Their effects can therefore be absorbed as small corrections 
to the condensate fields. This motivates the following ansatz: 
\begin{equation}
\label{ansatz}
\left\{
\begin{array}{c}
\Sigma_{S}^{\rm ChPT} (k_f,\rho) = \frac{1}{2} U(k_f,k_f) \\
\Sigma_{V}^{\rm ChPT} (k_f,\rho) = \frac{1}{2} U(k_f,k_f) \\
\Sigma_{TS}^{\rm ChPT}(k_f,\rho) = -\frac{1}{2} U_{I}(k_f,k_f) \delta\\
\Sigma_{TV}^{\rm ChPT}(k_f,\rho) = -\frac{1}{2} U_{I}(k_f,k_f) \delta\\
\end{array}
\right. \; ,
\end{equation}
in which the single-nucleon self-energy Eq. (\ref{sum}) is
equally divided between the vector Eq. (\ref{self1}) and 
the scalar Eq. (\ref{self3}) components. For the ChPT 
isoscalar and isovector self-energies \cite{Fri.04} a
polynomial fit up to order $k_f^6$ is performed: 
\begin{eqnarray}
   \label{expansion1}
   U(k_f; b_{3},b_{5},b_6) & = &
    ~\left[ c_{3} + 2b_{3} \right] \frac{k_f^3}{\Lambda^2} 
   + c_{4} ~ \frac{k_f^4}{\Lambda^3} 
   + \left[ c_{5} + {8\over 3}b_{5} \right] 
   \frac{k_f^5}{\Lambda^4} \nonumber\\
& ~ & 
      +~ \left[ c_{6}+ 3b_6\right] \frac{k_f^6}{\Lambda^5} 
   + \mathcal{O}(k_f^7)\; 
\end{eqnarray}
and
\begin{eqnarray}
   \label{expansion2}
   U_I(k_f,a_{3},a_{5},a_6, b_5) & = &
   ~\left[ d_{3} + 2a_{3} \right] \frac{k_f^3}{\Lambda^2} 
   \nonumber\\
& ~ & + ~d_{4} ~ \frac{k_f^4}{\Lambda^3} + 
   \left[ d_{5} + 2a_5 - {10\over 9} b_{5} \right] 
   \frac{k_f^5}{\Lambda^4} \nonumber\\
& ~ & 
      + ~\left[ d_{6} + 2 a_6 \right] 
   \frac{k_f^6}{\Lambda^5} 
   + \mathcal{O}(k_f^7)\; ,
\end{eqnarray}
where $b_i$ and  $a_i$ are the dimensionless short-distance 
regularization constants appearing  
in the counter terms (\ref{de}) and (\ref{das}). 
They subsume all unresolved short-range 
two- and three-body dynamics, plus possible higher order 
(four-loop etc.) effects not evaluated 
but contributing already at order $k_f^6$. 
The constants $c_i$ and $d_i$ are derived directly 
from the in-medium ChPT calculations of the finite 
(regularization-independent) parts of one- and 
two-pion-exchange contributions to the energy, 
with inclusion of $\Delta(1232)$ intermediate 
states. They are taken over unchanged from 
Ref.~\cite{Fri.04}. Note that the coupling 
strengths $g_i$ in the isoscalar 
single-particle potential (\ref{sppot}) 
are identified as 
\begin{eqnarray}
  g_3 & = & c_3 + 2b_3 \; .
   \nonumber\\
 g_4 & = & c_4 \; ,
  \nonumber\\
 g_5 & = & c_5 + {8\over 5} b_5 \; ,
  \nonumber\\
  g_6 & = & c_6 + 3b_6 \; . 
\end{eqnarray}
The counter term constants $b_i$ and $a_i$ are 
listed in Table~\ref{tab3}. The constants $c_i$ 
and $d_i$ are collected in Table~\ref{tabA1} at 
the reference scale $\Lambda = 2\pi f_\pi \simeq 0.58$ GeV. 

In order to determine the density-dependent 
couplings of the exchange correlation pieces 
generated by the point-coupling model, the ChPT 
self-energies are first re-expressed in terms of the baryon
density $\rho = 2k_f^3/3\pi^2 = \rho^p + \rho^n$ 
and the isovector density $\rho_3 = \rho^p - \rho^n$:
\begin{eqnarray}
  \Sigma_S^{\rm ChPT}(k_f,\rho) & = & (c_{s1} + c_{s2}\rho^{\frac{1}{3}}
  + c_{s3}\rho^{\frac{2}{3}}+ c_{s4} \rho) \,\rho
  \; ,\label{prho1} \\
  \Sigma_V^{\rm ChPT}(k_f,\rho) & = & (c_{v1} + c_{v2}\rho^{\frac{1}{3}}
  + c_{v3}\rho^{\frac{2}{3}}+ c_{v4} \rho)\, \rho
  \; ,\label{prho2}\\
  \Sigma_{TS}^{\rm ChPT}(k_f,\rho) &=& (c_{ts1} + c_{ts2}\rho^{\frac{1}{3}}
  + c_{ts3}\rho^{\frac{2}{3}}+ c_{ts4} \rho )\, \rho_{3} 
  \; ,\label{prho3}\\
  \Sigma_{TV}^{\rm ChPT}(k_f,\rho) &=& (c_{tv1} + c_{tv2}\rho^{\frac{1}{3}}
  + c_{tv3}\rho^{\frac{2}{3}}+ c_{tv4} \rho)\, \rho_{3}
  \label{prho4}\; ,
\end{eqnarray}
Next these self-energies are identified with the corresponding 
contributions to the 
potentials in the point coupling single-nucleon 
Dirac equation (\ref{self1})-(\ref{self5}):
\begin{eqnarray}
\Sigma_S^{\rm PC}(k_f,\rho) & = & G_S^{(\pi)}(\rho) \,\rho_S \; ,\\
\Sigma_V^{\rm PC}(k_f,\rho) & = & G_V^{(\pi)}(\rho)\, \rho +
\frac{1}{2} \left\{ \frac{\partial G_V^{(\pi)}(\rho)}{\partial \rho}
\rho^2 + \frac{\partial G_S^{(\pi)}(\rho)
}{\partial \rho} \rho_S^2 \right\} \; ,\\
\Sigma_{TS}^{\rm PC}(k_f,\rho) & = & G_{TS}^{(\pi)}(\rho) \,\rho_{S3} \; , \\
\Sigma_{TV}^{\rm PC}(k_f,\rho) & = & G_{TV}^{(\pi)}(\rho)\, \rho_{3} \; ,
\end{eqnarray}
The resulting expressions for the density-dependent couplings
of the pionic fluctuation terms read\footnote{
small differences between $\rho$ and $\rho_S$ at nuclear matter densities
are neglected here.}
\begin{eqnarray}
\label{GS}
  G_S^{(\pi)} (\rho)& = & c_{s1} + c_{s2} \rho^{\frac{1}{3}}
  + c_{s3} \rho^{\frac{2}{3}} + c_{s4} \rho \; ,\\
\label{GV}
  G_V^{(\pi)} (\rho)& = & {\bf \bar{c}_{v1}} + 
{\bf \bar{c}_{v2}} \rho^{\frac{1}{3}}
  + {\bf \bar{c}_{v3}} \rho^{\frac{2}{3}} + {\bf \bar{c}_{v4}} \rho \; ,\\
\label{GTS}
  G_{TS}^{(\pi)} (\rho)& = & c_{ts1} + c_{ts2} \rho^{\frac{1}{3}}
  + c_{ts3} \rho^{\frac{2}{3}} + c_{ts4} \rho \; ,\\
\label{GTV}
  G_{TV}^{(\pi)} (\rho)& = & c_{tv1} + c_{tv2} \rho^{\frac{1}{3}}
  + c_{tv3} \rho^{\frac{2}{3}} + c_{tV4} \rho \; ,
\end{eqnarray}
where the inclusion of the rearrangement term $\Sigma_R^0$ redefines
the isoscalar-vector coefficients:
\begin{equation}
\left\{ \begin{array}{ccl}
{\bf \bar{c}_{v1}} & = & c_{v1} \\
{\bf \bar{c}_{v2}} & = & 
\frac{1}{7}(6c_{v2}-c_{s2}) \\
{\bf \bar{c}_{v3}} & = & 
\frac{1}{4}(3c_{v3}-c_{s3}) \\
{\bf \bar{c}_{v4}} & = & 
\frac{1}{3}(2c_{v4}-c_{s4}) \; .
\end{array} \right. 
\label{bar-c}
\end{equation}
The coefficients of the expansion of the density-dependent 
couplings in powers of the baryon density 
(Eqs. (\ref{GS}) -- (\ref{bar-c})) are listed
in Table~\ref{tabA2}. 
%%%%%%%%%%%%%%%%%%%%%%%%%%%%%%%%%%%%%%%%%%%%%%%%%%%%%%%%%%%%%%%%%%%%%%%%%%%%
\begin{table}
\begin{center}
\caption{Coefficients of the expansions 
(\ref{expansion1}-\ref{expansion2})
up to order $k_f^6$ in the isoscalar and isovector channel. 
The reference scale is $\Lambda = 2\pi f_\pi \simeq 0.58$ GeV.}
\bigskip
\begin{tabular}{|l|c|c|c|}
\hline
$c_{3}$ &   3.01   & $d_3$   &   -2.80  \\
$c_{4}$ &   2.67   & $d_{4}$ &   -0.35  \\
$c_{5}$ &   2.17   & $d_{5}$ &   -8.88  \\
$c_{6}$ &  13.70   & $d_{6}$ &   11.50  \\

\hline
\end{tabular}
\label{tabA1}
\end{center}
\end{table}
%%%%%%%%%%%%%%%%%%%%%%%%%%%%%%%%%%%%%%%%%%%%%%%%%%%%%%%%%%%%%%%%%%%%%%%%%%%%%%

%%%%%%%%%%%%%%%%%%%%%%%%%%%%%%%%%%%%%%%%%%%%%%%%%%%%%%%%%%%%%%%%%%%%%%%%%%%%%%
\begin{table}
\begin{center}
\caption{Coefficients of the expansion of the density-dependent
couplings (\ref{GS}) -- (\ref{GTV})
in powers of the baryon density.  The input parameters 
in this example are those of the
ChPT calculation ~\cite{Fri.04} 
($b_3 = -3.05, a_3 = 2.16, b_5 = a_5 = 0, b_6 = - a_6 = 2.82$). }
\bigskip
\begin{tabular}{|l|c|l|c|c|c|}
\hline
$c_{s1} = c_{v1}$ & -2.65 fm$^{2}$  & $c_{ts1} = c_{tv1}$ 
&  1.31 fm$^{2}$ & ${\bf \bar{c}_{v1}}$ & -2.65 fm$^{2}$ \\
$c_{s2} = c_{v2}$ &  1.91 fm$^{3}$  & $c_{ts2} = c_{tv2}$ 
& -0.25 fm$^{3}$ & ${\bf \bar{c}_{v2}}$ &  1.36 fm$^{3}$ \\
$c_{s3} = c_{v3}$ &  1.29 fm$^{4}$  & $c_{ts3} = c_{tv3}$ 
& -5.29 fm$^{4}$ & ${\bf \bar{c}_{v3}}$ &  0.65 fm$^{4}$ \\
$c_{s4} = c_{v4}$ &  2.59 fm$^{5}$  & $c_{ts4} = c_{tv4}$ 
&  6.65 fm$^{5}$ & ${\bf \bar{c}_{v4}}$ &  0.43 fm$^{5}$ \\
\hline
\end{tabular}
\label{tabA2}
\end{center}
\end{table}
%%%%%%%%%%%%%%%%%%%%%%%%%%%%%%%%%%%%%%%%%%%%%%%%%%%%%%%%%%%%%%%%%%%%%%%%%%%%%

%%%%%%%%%%%%%%%%%%%%%%%%%%%%%%%%%%%%%%%%%%%%%%%%%%%%%%%%%%%%%%%%%%%%%%%%%%%%%
~\\ 
~\\ \newpage

%%%%%%%%%%%%%%%%%%%%%%%%%%%%%%%%%%%%%%%%%%%%%%%%%%%%%%%%%%%%%%%%%%%%%%%%%%%%%

%%%%%%%%%%%%%%%%%%%%%%%%%%%%%%%%%%%%%%%%%%%%%%%%%%%%%%%%%%%%%%%%%%%%%%%%%%%%%%

%%%%%%%%%%%%%%%%%%%%%%%%%%%%%%%%%%%%%%%%%%%%%%%%%%%%%%%%%%%%%%%%%%%%%%%%%%%%%
\newpage
%%%%%%%%%%%%%%%%%%%%%%%%%%%%%%%%%%%%%%%%%%%%%%%%%%%%%%%%%%%%%%%%%%%%%%%%%%%%%
%%%%%%%%%%%%%%%%%%%%%%%%%%%%%%%%%%%%%%%%%%%%%%%%%%%%%%%%%%%%%%%%%%%%%%%%%%%%%%
\newpage
%%%%%%%%%%%%%%%%%%%%%%%%%%%%%%%%%%%%%%%%%%%%%%%%%%%%%%%%%%%%%%%%%%%%%%%%%%%%%
%%%%%%%%%%%%%%%%%%%%%%%%%%%%%%%%%%%%%%%%%%%%%%%%%%%%%%%%%%%%%%%%%%%%%%%%%%%%%%
\newpage
%%%%%%%%%%%%%%%%%%%%%%%%%%%%%%%%%%%%%%%%%%%%%%%%%%%%%%%%%%%%%%%%%%%%%%%%%%%%%%
%%%%%%%%%%%%%%%%%%%%%%%%%%%%%%%%%%%%%%%%%%%%%%%%%%%%%%%%%%%%%%%%%%%%%%%%%%%%%%
\newpage
%%%%%%%%%%%%%%%%%%%%%%%%%%%%%%%%%%%%%%%%%%%%%%%%%%%%%%%%%%%%%%%%%%%%%%%%%%%%%%
%%%%%%%%%%%%%%%%%%%%%%%%%%%%%%%%%%%%%%%%%%%%%%%%%%%%%%%%%%%%%%%%%%%%%%%%%%%%%%
\newpage
%%%%%%%%%%%%%%%%%%%%%%%%%%%%%%%%%%%%%%%%%%%%%%%%%%%%%%%%%%%%%%%%%%%%%%%%%%%%%%
%%%%%%%%%%%%%%%%%%%%%%%%%%%%%%%%%%%%%%%%%%%%%%%%%%%%%%%%%%%%%%%%%%%%%%%%%%%%%%
\newpage
%%%%%%%%%%%%%%%%%%%%%%%%%%%%%%%%%%%%%%%%%%%%%%%%%%%%%%%%%%%%%%%%%%%%%%%%%%%%%%
%%%%%%%%%%%%%%%%%%%%%%%%%%%%%%%%%%%%%%%%%%%%%%%%%%%%%%%%%%%%%%%%%%%%%%%%%%%%%%
\newpage
%%%%%%%%%%%%%%%%%%%%%%%%%%%%%%%%%%%%%%%%%%%%%%%%%%%%%%%%%%%%%%%%%%%%%%%%%%%%%%
%%%%%%%%%%%%%%%%%%%%%%%%%%%%%%%%%%%%%%%%%%%%%%%%%%%%%%%%%%%%%%%%%%%%%%%%%%%%%%
\newpage
%%%%%%%%%%%%%%%%%%%%%%%%%%%%%%%%%%%%%%%%%%%%%%%%%%%%%%%%%%%%%%%%%%%%%%%%%%%%%%
%%%%%%%%%%%%%%%%%%%%%%%%%%%%%%%%%%%%%%%%%%%%%%%%%%%%%%%%%%%%%%%%%%%%%%%%%%%%%%
\newpage
%%%%%%%%%%%%%%%%%%%%%%%%%%%%%%%%%%%%%%%%%%%%%%%%%%%%%%%%%%%%%%%%%%%%%%%%%%%%%%
%%%%%%%%%%%%%%%%%%%%%%%%%%%%%%%%%%%%%%%%%%%%%%%%%%%%%%%%%%%%%%%%%%%%%%%%%%%%%%
\newpage
%%%%%%%%%%%%%%%%%%%%%%%%%%%%%%%%%%%%%%%%%%%%%%%%%%%%%%%%%%%%%%%%%%%%%%%%%%%%%%
%%%%%%%%%%%%%%%%%%%%%%%%%%%%%%%%%%%%%%%%%%%%%%%%%%%%%%%%%%%%%%%%%%%%%%%%%%%%%%

\end{document}